\documentclass[12pt]{extarticle}%
\usepackage[a4paper, margin=1in]{geometry}
\usepackage{eurosym}
\usepackage{setspace}
\usepackage[utf8]{inputenc}
\usepackage{natbib}
\usepackage{amsmath,amsfonts,amssymb,amsthm,enumitem}
\usepackage{graphicx}
\usepackage{rotating}
\usepackage{subcaption}
\usepackage{multirow}
\usepackage{amsthm}
\usepackage{pgf}
\usepackage{setspace}
\usepackage{hyperref}
\usepackage{bbm}
\usepackage{comment}
\usepackage{booktabs}
\usepackage{tikz,calc}
\usepackage{accents}
\usepackage{xcolor}
\usepackage{soul}
\usepackage{pgfplots}
\usepackage{graphicx}
\usepackage{caption}
\usepackage{pdflscape}
\usepackage{float}
\usepackage{array}
\usepackage{multirow}
\usepackage{needspace}
\usepackage{placeins}
\usepackage{amsmath}
\usepackage{amsfonts}
\usepackage{xcolor}
\usepackage{amssymb}%
\usepackage{graphicx} % Required for inserting images
\providecommand{\U}[1]{\protect\rule{.1in}{.1in}}

\usepackage{authblk}
\usepackage{natbib}

\begin{document}

\title{Leverage Dynamics in Product-Market Networks}

\author[1]{Stylianos Asimakopoulos}
\author[1]{Vasilis Sarafidis}

\affil[1]{Brunel University London}

\date{}
\maketitle

\begin{abstract}
We examine whether corporate leverage persistence reflects only firms' own adjustment frictions or also the past financing choices of economically connected firms. We combine annual firm fundamentals with time-varying text-based product-market networks and estimate a heterogeneous dynamic partial-adjustment model with interactive effects using defactored instrumental variables. The estimator accommodates time- varying network weights and spatial dependence in firm-specific coefficients. Network-weighted lagged leverage is positive and statistically significant, with a mean coefficient representing approximately 45\% of the sum of the own and network lag coefficients for book leverage. Accounting for network dependence reduces the persistence attributed to firms' own financing histories and increases the implied own-firm speed of adjustment. Firm-specific estimates reveal significant product-market clustering in own-leverage persistence and in responses to several leverage determinants, showing that heterogeneity in financing behaviour is systematically related to product-market proximity. The evolution of product-market connections matters quantitatively for network dependence: holding the network fixed produces a substantially larger network coefficient and considerably larger indirect effects than using annual network weights. Network dependence is therefore an economically important component of leverage dynamics, while its measurement depends on both the specification of economic connections and heterogeneity in firm behaviour.
\end{abstract}



\baselineskip 0.27in

\maketitle

\newpage

\section{Introduction}

Corporate leverage is highly persistent. In standard partial-adjustment models, this persistence is attributed to firms adjusting gradually towards target leverage because changing capital structure is costly \citep{FlanneryRangan2006,Dang2013ECM}. A related literature establishes that firms do not make financing decisions in isolation. \citet{LearyRoberts2014} show that firms respond to the financial policies of their peers, while \citet{GrieserEtAl2022} document sizeable peer effects within product-market networks. This raises a natural question: how much of observed leverage persistence reflects a firm's own past financing choices, and how much reflects the past financing choices of economically connected firms?

Existing evidence on peer effects does not directly address this question. In particular, \citet{GrieserEtAl2022} model \textit{contemporaneou}s dependence in firms' financial policies using annual product-market networks: leverage of firm $i$ in year $t$ depends on the leverage of economically connected firms in the same year. Their analysis establishes an important role for network interactions in corporate financial policy, but does not model these interactions as part of the adjustment of leverage from one period to the next. Conversely, conventional partial-adjustment models include the firm's own lagged leverage but do not allow past leverage of connected firms to affect subsequent adjustment. We bring these two approaches together by allowing current leverage to depend jointly on the firm's own lagged leverage and network-weighted lagged leverage.

Establishing network effects in corporate financial decisions poses a fundamental challenge. As emphasised by \citet{Manski1993}, similar outcomes among members of a group may arise from endogenous interactions, but also from correlated effects: firms may behave similarly because they share characteristics or are exposed to the same environment. Both possibilities are particularly relevant in product-market networks. First, firms sort endogenously into product markets. Economically connected firms may therefore have similar financing behaviour not because their financing choices affect one another, but because firms with similar characteristics also have similar adjustment speeds and responses to leverage determinants. If firms differ systematically in how they adjust leverage and respond to its determinants, and firms with similar responses tend to be economically connected, imposing common responses can confound this clustering with network dependence. Second, connected firms may respond to common unobserved shocks. Macroeconomic, financial, technological, or industry-wide shocks can generate co-movement in leverage even in the absence of transmission between firms, particularly when firms differ in their exposure to those shocks. In either case, cross-sectional co-movement in leverage can be mistakenly attributed to network effects.

Our empirical framework addresses these identification concerns by extending existing models of corporate peer effects along two dimensions that have not generally been accommodated jointly. First, we allow all model coefficients to vary across firms, including their responses to their own past leverage, the past leverage of economically connected firms, and other leverage determinants. Importantly, this heterogeneity may itself be clustered within the product-market network. Second, we allow the regression error to have a multifactor structure in which latent common shocks affect firms with heterogeneous factor loadings. Network dependence is therefore identified conditional on both firm-specific behavioural heterogeneity and heterogeneous exposure to common unobserved shocks.

To assess how much of observed leverage persistence reflects a firm's own past financing choices and how much reflects the past financing choices of economically connected firms, we use a balanced panel of 549 U.S. listed firms over 1990--2023. Product-market connections are measured using the text-based product similarity measures of \citet{HobergPhillips2010RFS,HobergPhillips2016JPE}. These measures provide firm-specific product-market weights and allow the strength of economic connections between firms to change from year to year. For each firm and year, we construct network-weighted leverage and fundamentals using the product-similarity weights prevailing at the relevant date. Our main specification asks whether the past leverage of economically connected firms helps explain a firm's current leverage, conditional on its own past leverage, observed fundamentals, firm-specific slope coefficients, and latent common shocks.

Estimating this specification requires accommodating dynamic adjustment, endogenous network variables, firm-specific coefficients, and latent common factors simultaneously. We therefore develop a mean-group instrumental-variable estimator based on defactored instruments, building on \citet{CuiSarafidisYamagata2023} and \citet{ChenCuiSarafidisYamagata2026}. We extend their framework in two directions. First, we allow the network weighting matrix to vary over time, so that the relevant economic connections can change from year to year. Second, we develop mean-group inference robust to spatial dependence in the firm-specific coefficients. The resulting estimator accommodates heterogeneous leverage dynamics and exposure to latent common shocks without imposing cross-sectional independence of firm-specific coefficients.

Our paper makes three main contributions. First, we contribute to the literature on dynamic capital structure adjustment by incorporating the past leverage of economically connected firms into the partial-adjustment framework. This allows persistence associated with a firm's own leverage history to be distinguished from dependence on past leverage elsewhere in the product-market network, with direct implications for the estimated speed at which firms adjust towards target leverage.

Second, we contribute to the identification of network effects in corporate financial decisions. By allowing firms to differ in their leverage dynamics and responses to fundamentals, and in their exposure to latent common shocks, we distinguish network dependence from correlated financing behaviour arising from endogenous sorting or exposure to a common unobserved economic environment. We also examine whether this behavioural heterogeneity is itself systematically related to product-market proximity.

Third, we make an econometric contribution by extending mean-group instrumental-variable estimation to heterogeneous dynamic network panels with time-varying network weights and spatially dependent firm-specific coefficients. The resulting framework accommodates dynamic endogeneity, coefficient heterogeneity, and latent common factors, and applies more generally to settings in which economic outcomes are linked through evolving observed networks.

We find strong evidence of lagged network dependence in corporate leverage. The mean coefficient on network-weighted lagged leverage is positive, statistically significant, and sizeable relative to the coefficient on the firm's own lagged leverage. Accounting for this network component substantially reduces the persistence attributed to the firm's own leverage history and correspondingly increases the implied own-firm speed of adjustment. Thus, conventional dynamic capital structure models that omit lagged network leverage can attribute to own-firm persistence dependence that is associated with past financing choices elsewhere in the product-market network.

The firm-specific estimates also provide evidence that heterogeneity in financing behaviour is systematically related to product-market proximity. To our knowledge, this is the first evidence that firm-specific capital-structure responses themselves exhibit spatial clustering within product-market networks. Firms that are closer in product-market space exhibit more similar degrees of own-leverage persistence and more similar responses to several leverage determinants, including tangibility, profitability, and firm size. This evidence supports allowing firm-specific responses when identifying network dependence.

The timing of product-market connections is also quantitatively important. Holding product-similarity weights fixed over the sample produces substantially stronger estimated network dependence than allowing the weights to vary annually. This distinction is specific to the dynamic setting studied here. Although \citet{GrieserEtAl2022} also construct annual product-market networks, their model relates firms' financial policies contemporaneously. In our setting, past financing choices enter the subsequent leverage-adjustment process, making the network prevailing when those choices were made relevant for measuring network dependence.

Our findings complement \citet{LearyRoberts2014} and \citet{GrieserEtAl2022}, who establish that the financial policies of economically related firms matter for firms' financing decisions. We show that this dependence extends across time: the past leverage of economically connected firms predicts subsequent leverage after controlling for the firm's own leverage history, heterogeneous responses to fundamentals, and latent common shocks. Product-market connections therefore enter not only firms' financing decisions but also the dynamic adjustment of leverage towards its target.

\section{Related Literature}

\subsection{Dynamic capital structure and partial adjustment}

The partial-adjustment model provides the standard dynamic interpretation of corporate leverage. Firms have a target leverage ratio but, because adjustment is costly, close only part of the gap between current and target leverage in each period. The coefficient on lagged leverage therefore measures the persistence of the firm's own capital structure and determines the implied speed of adjustment towards target leverage. \citet{FlanneryRangan2006} provide influential evidence of economically meaningful adjustment towards target leverage, while subsequent work shows that adjustment speeds vary substantially across firms and economic conditions \citep{Dang2013ECM}.

A separate literature identifies persistence in terms of the tendency of cross-sectional differences in leverage to remain pronounced over time. \citet{LemmonRobertsZender2008} show that firm-specific effects account for a substantial part of the variation in leverage and remain important over long periods. At the same time, \citet{DeAngeloRoll2015} document substantial changes in firms' capital structures over time, including changes in their relative leverage positions. These findings distinguish persistence in cross-sectional differences from persistence in an individual firm's leverage ratio: substantial differences across firms can persist even when individual firms experience large changes in leverage.

Our analysis is more directly related to the first notion of persistence, but broadens the conventional partial-adjustment model in two respects: leverage dynamics are allowed to differ across firms, and current leverage is allowed to depend on the past leverage of economically connected firms. The model therefore distinguishes persistence associated with a firm's own financing history from dependence on past financing choices elsewhere in its product-market network. This distinction matters directly for the estimated speed of adjustment: if lagged network leverage is omitted, part of the persistence attributed to the firm's own capital structure may instead reflect the past financing choices of connected firms.

\subsection{Network effects in corporate financial policy}

Corporate financing decisions, however, need not depend only on a firm's own characteristics and financing history. \citet{LearyRoberts2014} show that firms respond to the
financing decisions of their peers and provide evidence that these responses operate through financing actions rather than simply through observable peer characteristics. Using idiosyncratic equity shocks to peer firms as a source of exogenous variation, they find economically sizeable effects of peer leverage on firms' own leverage. Their results provide an important benchmark for our analysis: financing choices made by other firms can affect a firm's own capital structure.

A central difficulty in this literature is distinguishing such interactions from correlated behaviour. As formalised by \citet{Manski1993}, firms may make similar decisions because they affect one another, but also because firms with similar characteristics or financing behaviour tend to be economically connected or because they face common shocks. \citet{GrieserEtAl2022} study this problem using the heterogeneous and intransitive product-market relationships measured by \citet{HobergPhillips2010RFS,HobergPhillips2016JPE}. Their spatial autoregressive specification relates each firm's financial policy to the contemporaneous policies of firms to which it is connected, with the strength of each connection determined by product similarity. They find positive and economically meaningful interactions in leverage and other corporate financial policies.

Our paper builds on this literature but asks whether network dependence operates across time. Whereas \citet{GrieserEtAl2022} relate a firm's leverage to the contemporaneous leverage of firms connected to it through the product market, we examine whether network-weighted leverage in period $t-1$ affects the firm's leverage in period $t$, conditional on its own leverage in $t-1$. This incorporates network dependence directly into the partial-adjustment process and allows own-firm persistence to be distinguished from dependence on the past financing choices of economically connected firms.

We also depart from existing corporate network models in how we accommodate correlated financing behaviour. We allow the coefficients governing own leverage dynamics, network dependence, and responses to fundamentals to vary across firms, including the possibility that this heterogeneity is itself clustered within the product-market network. We also allow latent common shocks to affect firms with heterogeneous factor loadings. These features reduce the scope for behavioural heterogeneity or common unobserved shocks to be attributed to network dependence.

\subsection{Product-market networks and the structure of propagation}

The measurement of economic connections determines which firms enter the network and the strength of their links. Conventional industry classifications assign firms to discrete groups, implicitly treating firms within a group as equally relevant peers. Product-market relationships are considerably richer. \citet{HobergPhillips2010RFS} use firms' product descriptions to measure pairwise product similarity, while \citet{HobergPhillips2016JPE} develop text-based network industry classifications that allow competitive relationships to overlap and vary in strength across pairs of firms

This distinction is particularly important in a network model because the elements of the weighting matrix determine the contribution of each connected firm $j$ to firm $i$'s network exposure. Product similarity therefore determines both which firms are economically relevant to one another and their relative weights in the network. Two firms can consequently have quite different network exposures even when they belong to the same conventional industry.

\citet{GrieserEtAl2022} demonstrate the empirical importance of this network structure for corporate financial policies. Using the Hoberg--Phillips measures, they show that estimates of financial-policy interactions depend materially on how product-market connections are defined and weighted. Network construction therefore affects the estimated magnitude of interactions among firms, rather than merely determining which firms are labelled as peers.

Product-market weights also change over time as firms alter their product offerings and competitive positions. The annual Hoberg--Phillips measures therefore generate a sequence of weighting matrices, $W_t$, rather than a single network that applies throughout the sample. \citet{GrieserEtAl2022} also use annual product-market information, so time variation in $W_t$ is not itself a contribution of our paper.

Time variation in the network takes on a different role in our dynamic setting because past financing choices are transmitted through product-market connections prevailing at particular points in time. A fixed weighting matrix replaces these changing relationships with a common set of weights over the sample. Comparing fixed and annual weighting matrices therefore allows us to assess whether changes in the strength of product-market connections matter for the estimated network component of leverage persistence.

The weighting matrix also determines network dependence maps into heterogeneous exposure across firms. The effect associated with a change in one firm's financing choice depends on both the estimated network response and its position within the pattern of product-market connections. Network structure can therefore generate substantial heterogeneity in the effects associated with a common network coefficient. This motivates our analysis of firm-specific effects and connects the paper to the broader literature on economic propagation through networks. \citet{OzdagliWeber2026}, for example, show how production linkages shape the direct and network components of monetary-policy transmission.

Our framework combines dynamic leverage adjustment with product-market network dependence to examine how past financing choices of economically connected firms contribute to subsequent leverage dynamics.

%The three strands of literature therefore play distinct roles in our analysis. The dynamic capital-structure literature provides the partial-adjustment framework and the conventional interpretation of leverage persistence. The corporate network-effects literature establishes that firms' financial decisions can respond to the decisions of economically connected firms and highlights the challenge of distinguishing such responses from correlated behaviour. The product-market network literature determines how those economic connections are measured, how their strength varies across firms and over time, and how network structure shapes the propagation of financing choices. Our model combines these elements to study how past financing choices propagate through product-market connections and contribute to subsequent leverage persistence.

\section{Methodology}\label{sec:methodology}

\subsection{Dynamic capital structure adjustment with network effects}

We begin with the standard partial-adjustment model of capital structure. Let $y_{i,t}^{*}$ denote firm $i$'s target leverage ratio in period $t$, that is, the leverage ratio the firm would choose in the absence of adjustment frictions. Because changing capital structure is costly, the firm need not move immediately to this target. Instead, it closes a fraction $\vartheta_i$ of the gap between its leverage at the end of the previous period and its current target:
\begin{equation}
y_{i,t}-y_{i,t-1}
=
\vartheta_i
\left(
y_{i,t}^{*}-y_{i,t-1}
\right)
+\varepsilon_{i,t},
\qquad
0<\vartheta_i\leq1,
\label{eq:partial_adjustment}
\end{equation}
where $\vartheta_i$ is firm $i$'s speed of adjustment. When $\vartheta_i=1$, the firm adjusts fully to its target within one period;
when $\vartheta_i<1$, adjustment is only partial and past leverage therefore continues to affect current leverage. The disturbance $\varepsilon_{i,t}$ captures idiosyncratic innovations to
observed leverage.

We extend the standard partial-adjustment model by allowing the target leverage ratio itself to depend on the past financing choices of economically connected firms:
\begin{equation}
y_{i,t}^{*}
=
\kappa_i
\sum_{j=1}^{N}w_{ij,t-1}y_{j,t-1}
+
\mathbf{x}_{i,t}^{\prime}\boldsymbol{\delta}_i
+
\sum_{j=1}^{N}w_{ij,t}
\mathbf{x}_{j,t}^{\prime}\boldsymbol{\eta}_i
+
\boldsymbol{\ell}_{i}^{\prime}\mathbf f_t .
\label{eq:target_leverage}
\end{equation}
Here, $\kappa_i$ measures the sensitivity of firm $i$'s target leverage to the past leverage of economically connected firms. The remaining terms allow the target to depend on the firm's own observed characteristics, the characteristics of connected firms, and latent common factors.

Several mechanisms can motivate such dependence. Firms may use the capital structures of economically relevant firms as financing benchmarks, may learn from their financing choices about the costs and benefits of alternative capital structures, or may adjust their desired leverage in response to the financial policies of firms with which they compete in product markets
\citep{LearyRoberts2014,GrieserEtAl2022}. These mechanisms all imply that the financing choices of connected firms can affect the leverage ratio that a firm would like to attain, while adjustment costs determine how quickly it moves towards that target.

We do not attempt to distinguish among these channels. Instead, we represent their combined effect by allowing firm $i$'s target leverage to depend on network-weighted leverage observed in the preceding period. Using past rather than contemporaneous network leverage also establishes a clear temporal ordering: financing choices made by connected firms in period $t-1$ can affect firm $i$'s target leverage in period $t$. This timing removes the contemporaneous reflection between firms' leverage choices, although lagged network leverage may still be endogenous; this remaining endogeneity is addressed by the instrumental-variable strategy developed below.

Substituting equation~(\ref{eq:target_leverage}) into equation~(\ref{eq:partial_adjustment}) yields
\begin{equation}
y_{i,t}
=
\alpha_i y_{i,t-1}
+
\psi_i
\sum_{j=1}^{N}w_{ij,t-1}y_{j,t-1}
+
\mathbf{x}_{i,t}^{\prime}\boldsymbol{\beta}_i
+
\sum_{j=1}^{N}w_{ij,t}
\mathbf{x}_{j,t}^{\prime}\boldsymbol{\gamma}_i
+
u_{i,t},
\label{eq:model_scalar}
\end{equation}
where
\[
\alpha_i=1-\vartheta_i,\qquad
\psi_i=\vartheta_i\kappa_i,\qquad
\boldsymbol{\beta}_i=\vartheta_i\boldsymbol{\delta}_i,\qquad
\boldsymbol{\gamma}_i=\vartheta_i\boldsymbol{\eta}_i,\qquad
\boldsymbol{\lambda}_i=\vartheta_i\boldsymbol{\ell}_i,
\]
and
\[
u_{i,t}
=
\boldsymbol{\lambda}_{i}^{\prime}\mathbf{f}_{t}
+
\varepsilon_{i,t},
\]
for $i=1,\ldots,N$ and $t=1,\ldots,T$. The $K\times 1$ vector
$\mathbf{x}_{i,t}=(x_{1,i,t},\ldots,x_{K,i,t})^{\prime}$ contains observed firm-specific covariates, including cash holdings, market-to-book, tangibility, profitability, SG\&A expenses, capital expenditure, firm size, and dividend-paying status, while $\boldsymbol{\beta}_{i}=(\beta_{1,i},\ldots,\beta_{K,i})^{\prime}$ and $\boldsymbol{\gamma}_{i}=(\gamma_{1,i},\ldots,\gamma_{K,i})^{\prime}$ denote the heterogeneous coefficients on own-firm and network-weighted characteristics, respectively. The $N\times N$ matrix $\mathbf W_t=(w_{i,j,t})$ describes the product-market network at time $t$, with $w_{i,i,t}=0$, and is allowed to vary over time as product-market relationships change.

The specification in (\ref{eq:model_scalar}) gives the two dynamic coefficients distinct economic interpretations. The parameter
$1-\alpha_i=\vartheta_i$ is firm $i$'s speed of adjustment towards target leverage. By contrast, $\psi_i=\vartheta_i\kappa_i$ measures the effect of network-weighted past leverage on current leverage. It combines the sensitivity of the firm's target leverage to the past financing choices of connected firms, $\kappa_i$, with the speed at which the firm adjusts towards that target, $\vartheta_i$. Thus, the effect of connected firms' past financing choices on current leverage is attenuated by the firm's adjustment speed: only the fraction $\vartheta_i$ of their effect on target leverage is reflected in current leverage within the period. The network-weighted covariates have an analogous interpretation: $\eta_i$ captures their effect on target leverage, while $\gamma_i=\vartheta_i \eta_i$ captures their effect on current leverage.

\subsection{Identification and economic interpretation}
\label{subsec:identification}

Identification of the network effect requires distinguishing dependence on the past financing choices of economically connected firms from correlated financing behaviour arising from common shocks or from similarities among connected firms  \citep{Manski1993,LearyRoberts2014,GrieserEtAl2022}. Using lagged rather than contemporaneous network leverage removes the mechanical simultaneity between a firm's current leverage and the current leverage of connected firms, but does not by itself establish exogeneity. Persistent common shocks may affect both the firm and firms to which it is economically connected, while firms with similar characteristics or behavioural responses may sort into similar product markets.

We address these concerns in two ways. First, the model allows firms to load heterogeneously on latent common factors, so that correlated financing behaviour need not be attributed to the observed network. Second, the instrumental-variable procedure developed below uses variation in observed covariates after removing their common-factor components. Identification of the dynamic network coefficient therefore comes from idiosyncratic variation that predicts lagged own and
network-weighted leverage while remaining orthogonal to the  contemporaneous idiosyncratic disturbance under the maintained moment conditions.

\subsection{Econometric methodology}
\label{subsec:iv}

To develop the estimator, we first express equation~(\ref{eq:model_scalar})
in time-series form for each firm. We index the $T$ estimation periods by
$t=1,\ldots,T$, with $t=0$ denoting the presample observation used to
construct lagged variables. Stacking observations over $t$, define
\[
\mathbf y_i=(y_{i,1},\ldots,y_{i,T})^{\prime},
\qquad
\mathbf y_{i,-1}=(y_{i,0},\ldots,y_{i,T-1})^{\prime},
\]
and
\[
\mathbf X_i
=
\begin{pmatrix}
\mathbf x_{i,1}^{\prime}\\
\mathbf x_{i,2}^{\prime}\\
\vdots\\
\mathbf x_{i,T}^{\prime}
\end{pmatrix},
\qquad
\mathbf X_i^{W}
=
\begin{pmatrix}
(\mathbf x_{i,1}^{W})^{\prime}\\
(\mathbf x_{i,2}^{W})^{\prime}\\
\vdots\\
(\mathbf x_{i,T}^{W})^{\prime}
\end{pmatrix},
\]
where
\[
\mathbf x_{i,t}^{W}
=
\sum_{j=1}^{N}w_{ij,t}\mathbf x_{j,t}.
\]

For $s\geq1$, define the corresponding $s$-period lagged covariate matrices as
\[
\mathbf X_{i,-s}
=
\begin{pmatrix}
\mathbf x_{i,1-s}^{\prime}\\
\mathbf x_{i,2-s}^{\prime}\\
\vdots\\
\mathbf x_{i,T-s}^{\prime}
\end{pmatrix},
\qquad
\mathbf X_{i,-s}^{W}
=
\begin{pmatrix}
(\mathbf x_{i,1-s}^{W})^{\prime}\\
(\mathbf x_{i,2-s}^{W})^{\prime}\\
\vdots\\
(\mathbf x_{i,T-s}^{W})^{\prime}
\end{pmatrix},
\]
where
\[
\mathbf x_{i,t-s}^{W}
=
\sum_{j=1}^{N}w_{ij,t-s}\mathbf x_{j,t-s}.
\]

The network-weighted lagged dependent variable is defined as
\[
\mathbf y_{i,-1}^{W}
=
\begin{pmatrix}
\displaystyle\sum_{j=1}^{N}w_{ij,0}y_{j,0}\\[0.3em]
\displaystyle\sum_{j=1}^{N}w_{ij,1}y_{j,1}\\
\vdots\\
\displaystyle\sum_{j=1}^{N}w_{ij,T-1}y_{j,T-1}
\end{pmatrix}.
\]

\noindent
Therefore, the model in equation~(\ref{eq:model_scalar}) can then be expressed as
\begin{equation}
\mathbf y_i
=
\alpha_i\mathbf y_{i,-1}
+
\psi_i\mathbf y_{i,-1}^{W}
+
\mathbf X_i\boldsymbol{\beta}_i
+
\mathbf X_i^{W}\boldsymbol{\gamma}_i
+
\mathbf u_i,
\label{eq:model_vector}
\end{equation}
where
\[
\mathbf u_i
=
\mathbf F\boldsymbol{\lambda}_i+\boldsymbol{\varepsilon}_i,
\qquad
\mathbf F=(\mathbf f_1,\ldots,\mathbf f_T)^{\prime}.
\]

For notational convenience, define
\begin{equation}
\boldsymbol{\theta}_i
=
\left(
\alpha_i,\,
\psi_i,\,
\boldsymbol{\beta}_i^{\prime},\,
\boldsymbol{\gamma}_i^{\prime}
\right)^{\prime},
\qquad
\mathbf C_i
=
\left(
\mathbf y_{i,-1},\,
\mathbf y_{i,-1}^{W},\,
\mathbf X_i,\,
\mathbf X_i^{W}
\right).
\label{eq:theta_C}
\end{equation}
Equation~(\ref{eq:model_vector}) can therefore be written compactly as
\begin{equation}
\mathbf y_i
=
\mathbf C_i\boldsymbol{\theta}_i+\mathbf u_i.
\label{eq:compact_model}
\end{equation}

Estimation of (\ref{eq:compact_model}) must address the endogeneity of own and network-weighted lagged leverage and the dependence of the observed regressors on the latent common factors. We use an instrumental-variable procedure based on the idiosyncratic, or defactored, components of the observed covariates. Following the factor representation used in \citet{NorkuteEtAl2021} and \citet{CuiNorkuteSarafidisYamagata2022}, suppose that
\begin{equation}
\mathbf x_{i,t}
=
\boldsymbol{\Gamma}_i^{\prime}\mathbf f_t
+
\mathbf v_{i,t},
\label{eq:x_factor}
\end{equation}
where $\boldsymbol{\Gamma}_i$ is an $r\times K$ matrix of factor loadings and
$\mathbf v_{i,t}$ denotes the idiosyncratic component of the covariates.
Stacking over time gives
\begin{equation}
\mathbf X_i
=
\mathbf F\boldsymbol{\Gamma}_i+\mathbf V_i.
\label{eq:X_factor}
\end{equation}

\noindent
Let
\[
\mathbf F_{-1}
=
(\mathbf f_0,\ldots,\mathbf f_{T-1})^{\prime},
\]
and define the corresponding defactoring matrices
\begin{equation}
\mathbf M_{\mathbf F}
=
\mathbf I_T-
\mathbf F(\mathbf F^{\prime}\mathbf F)^{-1}\mathbf F^{\prime},
\qquad
\mathbf M_{\mathbf F_{-1}}
=
\mathbf I_T-
\mathbf F_{-1}
(\mathbf F_{-1}^{\prime}\mathbf F_{-1})^{-1}
\mathbf F_{-1}^{\prime}.
\label{eq:defactoring}
\end{equation}

For $s=0,1,\ldots$, define the $s$-period lagged network-weighted covariate
matrix as
\begin{equation}
\mathbf X_{i,-s}^{W}
=
\begin{pmatrix}
(\mathbf x_{i,1-s}^{W})^{\prime}\\
(\mathbf x_{i,2-s}^{W})^{\prime}\\
\vdots\\
(\mathbf x_{i,T-s}^{W})^{\prime}
\end{pmatrix},
\qquad
\mathbf x_{i,t-s}^{W}
=
\sum_{j=1}^{N}w_{ij,t-s}\mathbf x_{j,t-s}.
\label{eq:lagged_network_x}
\end{equation}

The infeasible instrument matrix for the regressors in $\mathbf C_i$ is
\begin{equation}
\mathbf Z_i
=
\left(
\mathbf M_{\mathbf F}\mathbf M_{\mathbf F_{-1}}\mathbf X_{i,-1},
\quad
\mathbf M_{\mathbf F}\mathbf M_{\mathbf F_{-1}}\mathbf X_{i,-1}^{W},
\quad
\mathbf M_{\mathbf F}\mathbf X_i,
\quad
\mathbf M_{\mathbf F}\mathbf X_i^{W}
\right).
\label{eq:population_Z}
\end{equation}
The first two blocks provide instruments for own and network-weighted lagged leverage, respectively, while the remaining blocks instrument the own and network-weighted contemporaneous covariates. The lagged defactored covariates retain idiosyncratic variation that is informative about lagged leverage while being orthogonal to the contemporaneous idiosyncratic disturbance under the maintained moment conditions. Similarly, the network-weighted defactored covariates provide variation that is informative about network-weighted leverage. The use of the time-varying network in $\mathbf X_{i,-1}^{W}$ ensures that these instruments reflect the network structure prevailing at the same date as the corresponding network-weighted lagged outcome.

Since $\mathbf F$ and $\mathbf F_{-1}$ are unobserved, they are estimated by principal components. Under the normalization
$T^{-1}\mathbf F^{\prime}\mathbf F=\mathbf I_r$, let
$\widehat{\mathbf F}$ be $\sqrt{T}$ times the eigenvectors associated with the $r$ largest eigenvalues of
\[
\frac{1}{NT}\sum_{i=1}^{N}\mathbf X_i\mathbf X_i^{\prime}.
\]
Similarly, let $\widehat{\mathbf F}_{-1}$ be $\sqrt{T}$ times the
eigenvectors associated with the $r$ largest eigenvalues of
\[
\frac{1}{NT}\sum_{i=1}^{N}
\mathbf X_{i,-1}\mathbf X_{i,-1}^{\prime}.
\]
Define
\[
\mathbf M_{\widehat{\mathbf F}}
=
\mathbf I_T-
\widehat{\mathbf F}
(\widehat{\mathbf F}^{\prime}\widehat{\mathbf F})^{-1}
\widehat{\mathbf F}^{\prime},
\]
with $\mathbf M_{\widehat{\mathbf F}_{-1}}$ defined analogously. The feasible instrument matrix is
\begin{equation}
\widehat{\mathbf Z}_i
=
\left(
\mathbf M_{\widehat{\mathbf F}}
\mathbf M_{\widehat{\mathbf F}_{-1}}\mathbf X_{i,-1},
\quad
\mathbf M_{\widehat{\mathbf F}}
\mathbf M_{\widehat{\mathbf F}_{-1}}\mathbf X_{i,-1}^{W},
\quad
\mathbf M_{\widehat{\mathbf F}}\mathbf X_i,
\quad
\mathbf M_{\widehat{\mathbf F}}\mathbf X_i^{W}
\right).
\label{eq:feasible_Z}
\end{equation}

Let
\begin{equation}
\widehat{\mathbf A}_i
=
T^{-1}\widehat{\mathbf Z}_i^{\prime}\mathbf C_i,
\qquad
\widehat{\mathbf B}_i
=
T^{-1}\widehat{\mathbf Z}_i^{\prime}\widehat{\mathbf Z}_i,
\qquad
\widehat{\mathbf b}_{y,i}
=
T^{-1}\widehat{\mathbf Z}_i^{\prime}\mathbf y_i.
\label{eq:moment_matrices}
\end{equation}
The individual-specific IV estimator is then given by
\begin{equation}
\widehat{\boldsymbol{\theta}}_i
=
\left(
\widehat{\mathbf A}_i^{\prime}
\widehat{\mathbf B}_i^{-1}
\widehat{\mathbf A}_i
\right)^{-1}
\widehat{\mathbf A}_i^{\prime}
\widehat{\mathbf B}_i^{-1}
\widehat{\mathbf b}_{y,i}.
\label{eq:theta_hat}
\end{equation}

Estimation is conducted separately for each firm, so no common-coefficient restriction is imposed on $\boldsymbol{\theta}_i$. Although the individual estimates are informative about firm-specific responses, our primary estimands are the population mean coefficients
\[
\boldsymbol{\theta}=E(\boldsymbol{\theta}_i).
\]
Following \citet{ChenCuiSarafidisYamagata2026}, we estimate these means using the Mean Group Instrumental Variables (MGIV) estimator
\begin{equation}
\widehat{\boldsymbol{\theta}}_{MG}
=
\frac{1}{N}
\sum_{i=1}^{N}
\widehat{\boldsymbol{\theta}}_i.
\label{eq:MGIV}
\end{equation}

Inference for the population mean coefficients must account for the possibility
that the firm-specific coefficients are correlated across economically proximate
firms. Let
\[
\widehat{\mathbf e}_i
=
\widehat{\boldsymbol{\theta}}_i
-
\widehat{\boldsymbol{\theta}}_{MG}
\]
denote the estimated deviation of firm $i$'s coefficient vector from the
cross-sectional mean. We estimate the long-run cross-sectional covariance
matrix using the spatial HAC estimator
\begin{equation}
\widehat{\boldsymbol{\Omega}}_{\theta}
=
\frac{1}{N}
\sum_{i=1}^{N}
\sum_{j=1}^{N}
K\left(
\frac{d_{ij,N}}{b_N}
\right)
\widehat{\mathbf e}_i
\widehat{\mathbf e}_j^{\prime},
\label{eq:MG_cov_estimation}
\end{equation}
where $d_{ij,N}$ is a measure of economic distance between firms $i$ and $j$,
$K(\cdot)$ is a bounded symmetric kernel, and $b_N$ is a bandwidth parameter.
The estimated covariance matrix of
$\widehat{\boldsymbol{\theta}}_{MG}$ is then
\begin{equation}
\widehat{\operatorname{Var}}
\left(
\widehat{\boldsymbol{\theta}}_{MG}
\right)
=
\frac{1}{N}
\widehat{\boldsymbol{\Omega}}_{\theta},
\label{eq:MG_var_estimation}
\end{equation}
and standard errors are obtained from the square roots of its diagonal
elements.

This formulation allows the heterogeneous coefficients themselves to exhibit spatial dependence, so that firms close in economic space may have systematically similar responses. When the firm-specific coefficient deviations are cross-sectionally independent, the off-diagonal covariance terms vanish and the conventional Mean Group covariance estimator is obtained as a special case. The large-$N$, large-$T$ conditions under which $\widehat{\boldsymbol{\theta}}_{MG}$ is consistent and asymptotically normal, and under which $\widehat{\boldsymbol{\Omega}}_{\theta}$ consistently estimates the long-run cross-sectional covariance matrix of the heterogeneous coefficients,
are provided below.

Inference for individual-specific coefficients is based on the estimated covariance matrix of $\widehat{\boldsymbol{\theta}}_i$, whose form is given in Section~\ref{subsec:asymptotic}.

\subsection{Assumptions}
\label{subsec:assumptions}

The following assumptions provide sufficient conditions for consistency and
asymptotic normality of the individual-specific and MGIV estimators.

\medskip
\noindent
\textbf{Assumption A (Idiosyncratic errors).}
The idiosyncratic errors $\varepsilon_{i,t}$ are independently distributed
across $i$ and $t$, with
\[
E(\varepsilon_{i,t})=0,
\qquad
E(\varepsilon_{i,t}^{2})=\sigma_{\varepsilon}^{2}>0,
\qquad
\sup_{i,t}E|\varepsilon_{i,t}|^{8+\delta}<C
\]
for some $\delta>0$.

\medskip
\noindent
\textbf{Assumption B (Idiosyncratic components of the covariates).}
The process $\mathbf v_{i,t}$ is independent of $\varepsilon_{j,s}$ for all
$i,j,t,s$, satisfies
\[
E(\mathbf v_{i,t})=\mathbf 0,
\qquad
\sup_{i,t}E\|\mathbf v_{i,t}\|^{8+\delta}<C,
\]
and is stationary with weak cross-sectional and temporal dependence and
absolutely summable autocovariances.

\medskip
\noindent
\textbf{Assumption C (Common factors).}
The common factors $\mathbf f_t$ are independent of
$\mathbf v_{i,s}$ and $\varepsilon_{i,s}$ for all $i,s,t$,
are stationary with absolutely summable autocovariances, and satisfy
\[
E\|\mathbf f_t\|^{4}<C.
\]

\medskip
\noindent
\textbf{Assumption D (Factor loadings).}
The factor loadings $\boldsymbol{\Gamma}_i$ and
$\boldsymbol{\lambda}_i$ have uniformly bounded fourth moments and are
independent of $\mathbf v_{j,t}$, $\varepsilon_{j,t}$, and $\mathbf f_t$.
Moreover,
\[
\frac{1}{N}\sum_{i=1}^{N}
\boldsymbol{\Gamma}_i\boldsymbol{\Gamma}_i^{\prime}
\xrightarrow{p}
\boldsymbol{\Sigma}_{\Gamma},
\]
where $\boldsymbol{\Sigma}_{\Gamma}$ is positive definite.

\medskip
\noindent
\textbf{Assumption E (Time-varying network).}
For every $t$, the observed network matrix
$\mathbf W_t=(w_{ij,t})_{N\times N}$ satisfies:
\begin{enumerate}[label=(\roman*)]
    \item $w_{ii,t}=0$ for all $i$ and $t$;
    \item $\sup_t\|\mathbf W_t\|_{\infty}<C$ and
          $\sup_t\|\mathbf W_t\|_{1}<C$;
    \item $\mathbf W_t$ is predetermined with respect to
          $\{\varepsilon_{i,s}:s\geq t\}$ and is independent of the
          contemporaneous idiosyncratic innovations;
    \item the sequence $\{\mathbf W_t\}$ is uniformly bounded and satisfies
          the regularity conditions required for the relevant time averages
          involving $\mathbf W_t$ to possess finite probability limits.
\end{enumerate}

\medskip
\noindent
\textbf{Assumption F (Random coefficients and spatial dependence).}
The heterogeneous coefficient vector satisfies
\begin{equation}
\boldsymbol{\theta}_i
=
\boldsymbol{\theta}
+
\mathbf e_i,
\qquad
E(\mathbf e_i)=\mathbf 0,
\qquad
\sup_i E\|\mathbf e_i\|^{4+\delta}<C,
\label{eq:random_coeff}
\end{equation}
for some $\delta>0$. The random coefficient deviations $\mathbf e_i$ are
independent of the common factors and idiosyncratic innovations, but may be
weakly dependent across firms. Cross-sectional dependence declines sufficiently
rapidly with the economic distance $d_{ij,N}$ such that
\begin{equation}
\boldsymbol{\Omega}_{\theta}
=
\lim_{N\rightarrow\infty}
\frac{1}{N}
\sum_{i=1}^{N}
\sum_{j=1}^{N}
E\left(
\mathbf e_i\mathbf e_j^{\prime}
\right)
\label{eq:omega_theta}
\end{equation}
exists, is finite, and is positive definite, and
\begin{equation}
\frac{1}{\sqrt{N}}
\sum_{i=1}^{N}
\mathbf e_i
\xrightarrow{d}
N\left(
\mathbf 0,
\boldsymbol{\Omega}_{\theta}
\right).
\label{eq:rc_clt}
\end{equation}
The kernel $K(\cdot)$ and bandwidth $b_N$ used in
(\ref{eq:MG_cov_estimation}) satisfy the regularity conditions required for
consistent estimation of $\boldsymbol{\Omega}_{\theta}$.

\medskip
\noindent
\textbf{Assumption G (Dynamic stability).}
The heterogeneous dynamic coefficients and the sequence of time-varying
network matrices $\{\mathbf W_t\}$ satisfy
\begin{equation}
\sup_i|\alpha_i|
+
\sup_i|\psi_i|\,
\sup_t\|\mathbf W_t\|_{\infty}
<1
\qquad \text{a.s.}
\label{eq:stability}
\end{equation}

\medskip
\noindent
\textbf{Assumption H (Identification).}
For each $i$, define
\[
\mathbf A_i
=
T^{-1}\mathbf Z_i^{\prime}\mathbf C_i,
\qquad
\mathbf B_i
=
T^{-1}\mathbf Z_i^{\prime}\mathbf Z_i,
\]
and
\[
\boldsymbol{\Phi}_i
=
T^{-1}
E\left(
\mathbf Z_i^{\prime}
\boldsymbol{\varepsilon}_i
\boldsymbol{\varepsilon}_i^{\prime}
\mathbf Z_i
\right).
\]
The following conditions hold:
\begin{enumerate}[label=(\roman*)]
    \item $\mathbf A_i$ has full column rank and $\mathbf B_i$ is positive
          definite for all $i$;
    \item $\mathbf A_i$ and $\mathbf B_i$ have uniformly bounded
          $(2+2\delta)$-th moments;
    \item $\|\boldsymbol{\Phi}_i\|<C$ uniformly in $i$ and $T$;
    \item there exist finite full-rank matrices
          $\mathbf A_{i,0}$, positive definite matrices
          $\mathbf B_{i,0}$ and $\boldsymbol{\Phi}_{i,0}$ such that
          \[
          \mathbf A_i\xrightarrow{p}\mathbf A_{i,0},
          \qquad
          \mathbf B_i\xrightarrow{p}\mathbf B_{i,0},
          \qquad
          \boldsymbol{\Phi}_i\rightarrow\boldsymbol{\Phi}_{i,0}.
          \]
\end{enumerate}

\subsection{Asymptotic properties}
\label{subsec:asymptotic}

We establish the large-$N$, large-$T$ properties of the individual-specific IV estimator and the MGIV estimator.

\medskip
\noindent
\textbf{Theorem 3.1.}
Suppose Assumptions A--E, G and H(i)--H(iii) hold. Then, for each $i$, the individual-specific IV estimator defined in (\ref{eq:theta_hat}) is consistent:
\[
\widehat{\boldsymbol{\theta}}_i
\xrightarrow{p}
\boldsymbol{\theta}_i,
\qquad
N,T\rightarrow\infty.
\]
If, in addition, Assumption H(iv) holds and
$T/N^{2}\rightarrow0$, then
\begin{equation}
\sqrt{T}
\left(
\widehat{\boldsymbol{\theta}}_i-\boldsymbol{\theta}_i
\right)
\xrightarrow{d}
N\left(
\mathbf 0,\boldsymbol{\Sigma}_{i,0}
\right),
\label{eq:individual_AN}
\end{equation}
where
\begin{align}
\boldsymbol{\Sigma}_{i,0}
={}&
\left(
\mathbf A_{i,0}^{\prime}
\mathbf B_{i,0}^{-1}
\mathbf A_{i,0}
\right)^{-1}
\mathbf A_{i,0}^{\prime}
\mathbf B_{i,0}^{-1}
\boldsymbol{\Phi}_{i,0}
\mathbf B_{i,0}^{-1}
\mathbf A_{i,0}
\nonumber\\
&\times
\left(
\mathbf A_{i,0}^{\prime}
\mathbf B_{i,0}^{-1}
\mathbf A_{i,0}
\right)^{-1}.
\label{eq:individual_var}
\end{align}

\noindent
\textit{Proof. See Appendix.}

\medskip
\noindent
The asymptotic covariance matrix can be estimated by
\begin{align}
\widehat{\boldsymbol{\Sigma}}_i
={}&
\left(
\widehat{\mathbf A}_i^{\prime}
\widehat{\mathbf B}_i^{-1}
\widehat{\mathbf A}_i
\right)^{-1}
\widehat{\mathbf A}_i^{\prime}
\widehat{\mathbf B}_i^{-1}
\widehat{\boldsymbol{\Phi}}_i
\widehat{\mathbf B}_i^{-1}
\widehat{\mathbf A}_i
\nonumber\\
&\times
\left(
\widehat{\mathbf A}_i^{\prime}
\widehat{\mathbf B}_i^{-1}
\widehat{\mathbf A}_i
\right)^{-1},
\end{align}
where
\begin{equation}
\widehat{\boldsymbol{\Phi}}_i
=
\frac{1}{T}
\sum_{t=1}^{T}
\widehat{\mathbf z}_{i,t}
\widehat{\mathbf z}_{i,t}^{\prime}
\widehat u_{i,t}^{\,2}
=
T^{-1}
\widehat{\mathbf Z}_i^{\prime}
\operatorname{diag}
\left(
\widehat u_{i,1}^{2},\ldots,\widehat u_{i,T}^{2}
\right)
\widehat{\mathbf Z}_i,
\end{equation}
and
\[
\widehat{\mathbf u}_i
=
\mathbf y_i-\mathbf C_i\widehat{\boldsymbol{\theta}}_i.
\]

\medskip
\noindent
\textbf{Theorem 3.2.}
Suppose Assumptions A--H hold and
$N/T^{2}\rightarrow0$ as $N,T\rightarrow\infty$. Then the MGIV estimator
defined in (\ref{eq:MGIV}) is consistent for the population mean
$\boldsymbol{\theta}$:
\[
\widehat{\boldsymbol{\theta}}_{MG}
\xrightarrow{p}
\boldsymbol{\theta}.
\]
If, in addition, $N/T^{6/5}\rightarrow0$, then
\begin{equation}
\sqrt{N}
\left(
\widehat{\boldsymbol{\theta}}_{MG}
-
\boldsymbol{\theta}
\right)
\xrightarrow{d}
N\left(
\mathbf 0,\boldsymbol{\Omega}_{\theta}
\right),
\label{eq:MG_AN}
\end{equation}
where
\begin{equation}
\boldsymbol{\Omega}_{\theta}
=
\lim_{N\rightarrow\infty}
\frac{1}{N}
\sum_{i=1}^{N}
\sum_{j=1}^{N}
E\left(
\mathbf e_i\mathbf e_j^{\prime}
\right).
\label{eq:MG_long_run_var}
\end{equation}
Furthermore, the spatial HAC estimator
$\widehat{\boldsymbol{\Omega}}_{\theta}$ defined in
(\ref{eq:MG_cov_estimation}) is consistent:
\[
\widehat{\boldsymbol{\Omega}}_{\theta}
\xrightarrow{p}
\boldsymbol{\Omega}_{\theta}.
\]

\noindent
\textit{Proof. See Appendix.}

\section{Data}\label{sec:data}

\subsection{Sample construction and firm-level variables}

Our empirical analysis combines accounting and market information from the CRSP--Compustat Merged database with annual text-based product-market similarity data from the Hoberg--Phillips product-market data library \citep{HobergPhillips2010RFS,HobergPhillips2016JPE}. The sample covers U.S. publicly traded firms over 1990--2023 and contains 549 firms observed throughout the sample period, yielding 18,666 firm-year observations before accounting for the lags required by the dynamic and instrumental-variable specifications.

The choice and construction of the firm-level variables follow the capital structure and peer-effects literature, particularly 
\citet{LearyRoberts2014} and \citet{GrieserEtAl2022}. Our primary dependent variable is book leverage ($bl$), with market leverage ($ml$) used as an alternative measure in the robustness analysis. The explanatory variables capture firm characteristics commonly associated with capital structure: cash holdings, growth opportunities, asset tangibility, profitability, selling, general and administrative expenses, capital expenditure, firm size,
and dividend-paying status.

Cash holdings are cash and short-term investments divided by total assets; the market-to-book ratio captures growth opportunities; tangibility is net property, plant and equipment over total assets; and profitability is operating profitability divided by total assets. Selling, general and administrative expenses are divided by sales, while capital expenditure is divided by total assets. Firm size is the natural logarithm of total assets, and the dividend indicator equals one for dividend-paying firms. Exact Compustat definitions
are reported in the notes to  Table~\ref{tab:summary_stats}.

\subsection{Product market network}

We identify economically relevant peer relationships using the text-based
product market similarity data developed by
\citet{HobergPhillips2010RFS,HobergPhillips2016JPE}. The measures are based on firms' product descriptions and provide annual pairwise measures of product similarity. Unlike SIC- or NAICS-based classifications, they allow the strength of product-market connections to vary continuously across firm pairs and over time as firms change their product scope and competitive positioning.

Let $s_{i,j,t}$ denote the Hoberg--Phillips product-similarity score between firms $i$ and $j$ in year $t$. We retain all observed links and construct the annual weighting matrix $\mathbf W_t=(w_{i,j,t})$ as
\begin{equation}
w_{i,j,t}
=
\frac{s_{i,j,t}}
{\sum_{k\neq i}s_{i,k,t}},
\qquad i\neq j,
\label{eq:network_weights}
\end{equation}
with $w_{i,i,t}=0$. Each non-empty row of $\mathbf W_t$ therefore sums to one, so that $w_{i,j,t}$ measures the relative importance of firm $j$ in firm $i$'s product-market network.

The weighting matrix is updated annually. Network-weighted variables are constructed using the matrix corresponding to the relevant year. In particular, the dynamic network term is $\mathbf W_{t-1}\mathbf y_{t-1}$, so that connected firms' lagged leverage is weighted according to the product-market relationships prevailing when those financing choices are observed.

This construction builds on the product-market network approach of \citet{GrieserEtAl2022}, while introducing the network into a dynamic partial-adjustment model. The distinction is important for our analysis: rather than relating a firm's leverage to the contemporaneous leverage of connected firms, we examine whether their past financing choices affect its current leverage.

For comparison, we also construct a fixed weighting matrix by averaging each firm-pair weight over the sample period. The robustness analysis considers additional group-linear and industry-linear weighting matrices, constructed using the corresponding group and industry classifications.

\subsection{Descriptive statistics}

Table~\ref{tab:summary_stats} reports descriptive statistics for the variables used in the empirical analysis. Mean book leverage is 0.234, compared with 0.168 for market leverage. Book leverage exhibits substantial variation both across firms and over time, with between- and within-firm standard deviations
of 0.138 and 0.115, respectively. The corresponding figures for market leverage are 0.113 and 0.097. Average book leverage is similar to that reported in related U.S. capital-structure studies \citep{LearyRoberts2014,GrieserEtAl2022}.

The firm characteristics also display considerable variation. Cash holdings average 12.7\% of total assets, tangibility 28.7\%, and profitability 13.6\%, while 71.8\% of firm-year observations are associated with dividend-paying firms. Importantly for our analysis, the explanatory variables exhibit
meaningful within-firm variation as well as substantial cross-sectional dispersion. This is useful for identifying changes in leverage associated with firms' own characteristics and those of their product-market connections.

Table~\ref{tab:network_variation} documents variation in the product-market network over time. The left panel measures, for a given pair of firms $(i,j)$, how much their product similarity changes over the 34-year sample. The right panel measures how similar the entire network in year $t$ is to that in year $t-1$. 

Pairwise connections exhibit substantial time-series variation. Average product similarity is 0.014, compared with an average within-pair standard deviation of 0.012, and the median within-pair range is 0.037. The median coefficient of variation is 1.617, implying that, for the median pair for which it is defined, the time-series standard deviation of product similarity is about
1.6 times its mean. In other words, the strength of a typical pairwise product-market connection varies substantially around its long-run average, rather than remaining close to a fixed level over time.

At the same time, the network is persistent from one year to the next, with an average correlation of 0.891 between consecutive annual similarity matrices. Despite this persistence, the mean absolute annual change in pairwise similarity is 0.006. This is economically sizeable relative to the average similarity of 0.014, amounting to approximately 44\% of its average level. Changes are also substantial for a non-negligible fraction of firm pairs. In an average year, about 15\% of pairs experience an absolute change in similarity exceeding one-half of the overall standard deviation of pairwise similarities. Thus,
although the broad structure of the product-market network is highly persistent, the strength of individual connections exhibits substantial year-to-year variation, motivating the use of annual network weights in the empirical analysis.

\section{Empirical results}

\subsection{Dynamic leverage adjustment and network effects}

Table~\ref{tab:baseline_network} reports our main estimates of the slope coefficients. We begin with the last column (column (7)), which reports S--MGIV under the time-varying network and corresponds to the model developed in Section~\ref{sec:methodology}. The specification allows the coefficients governing leverage adjustment, network dependence, and responses to firm characteristics to differ across firms; accommodates latent common shocks with heterogeneous factor loadings; instruments the endogenous dynamic terms; and allows product-market weights to vary annually. Moving from right to left, the preceding columns impose progressively stronger restrictions and allow us to isolate the importance of coefficient heterogeneity, network dependence, and time variation in product-market connections.

A key finding is that current leverage depends strongly on both the firm's own leverage history and the past leverage of economically connected firms. The mean coefficient on own lagged leverage is 0.346 (SE = 0.013), while the mean coefficient on network-weighted lagged leverage is 0.283 (SE = 0.072); both are significant at the 1\% level. Under the partial-adjustment interpretation developed in Section~\ref{sec:methodology}, the own-lag coefficient implies an average speed of adjustment of $1-0.346=0.654$, so firms close  approximately 65\% of the gap between lagged and target leverage within one year. This indicates relatively rapid adjustment towards target leverage and is consistent with evidence that firms make economically meaningful adjustments towards their capital-structure targets \citep{FlanneryRangan2006,Dang2013ECM}. Importantly, however, our adjustment speed is estimated conditional on the past leverage of economically connected firms. It therefore measures the rate of own-firm adjustment after separating out persistence associated with financing choices elsewhere in the product-market network. 

The network coefficient has a different interpretation. Since $\psi_i=\vartheta_i\kappa_i$, it measures the effect of connected firms' past leverage on current observed leverage, combining the sensitivity of the firm's target leverage to those financing choices with the speed at which the firm moves towards its target. The estimate of 0.283 therefore identifies a sizeable
dynamic network component alongside firms' own leverage persistence. Its magnitude is also economically substantial relative to existing evidence on financing interactions. \citet{LearyRoberts2014} document sizeable responses of firms' leverage to the financing decisions of industry peers, while \citet{GrieserEtAl2022} report leverage interaction coefficients of approximately
0.13--0.23 across alternative product-market network specifications. Our estimate of 0.283 is somewhat larger, although the parameters are not directly
comparable. In particular, \citet{GrieserEtAl2022} relate current leverage to the contemporaneous leverage of connected firms, whereas our coefficient measures the response of current leverage to their leverage in the preceding year, conditional on the firm's own lagged leverage. The comparison nevertheless shows that the dynamic network dependence identified here is economically meaningful relative to previously documented financing interactions.

The remaining estimates in the last column describe how own-firm and network-level fundamentals enter the adjustment process. Cash and profitability are negatively associated with leverage, with mean coefficients of $-0.107$ and $-0.429$, respectively, consistent with firms relying less on debt when internal liquidity and internally generated funds are greater. Tangibility enters positively (0.130), consistent with tangible assets supporting debt capacity through their collateral value, while the positive coefficient on size (0.055) accords with the tendency of larger firms to sustain higher leverage. Market-to-book is also positive and statistically significant, whereas SG\&A, capital expenditure, and the dividend indicator are not statistically significant.

The network-weighted characteristics control for the economic conditions of connected firms. Network-weighted size enters negatively ($-0.022$, p$<0.01$), network-weighted market-to-book is negative at the 10\% level, and network-weighted profitability is not statistically significant. Their inclusion ensures that the dynamic network coefficient is estimated conditional on observable characteristics of connected firms.\footnote{We do not present results with all firm characteristics included in network-weighted form. The additional network-weighted covariates are largely
statistically insignificant, with $p$-values generally above 0.7, while their inclusion reduces the precision of the remaining coefficient estimates.}

We next use columns (1)--(6) to assess how the estimated leverage dynamics change when key features of the full model in column (7) are restricted. We first examine what is attributed to own-firm persistence when network dependence is omitted. MGIV provides the natural benchmark because it retains coefficient heterogeneity
and the multifactor error structure but excludes the network terms. The mean coefficient on own lagged leverage is 0.462 under MGIV, compared with 0.346
under S--MGIV with the time-varying network. The corresponding speed of adjustment increases from 53.8\% to 65.4\%, a difference of 11.6 percentage points. Thus, when network dependence is omitted, substantially more of the observed persistence in leverage is attributed to the firm's own leverage history. The relative magnitude of the two dynamic components further illustrates this point: network-weighted lagged leverage accounts for
approximately 45\% of the sum of the mean own- and network-lag coefficients under the full model in column (7). This is a descriptive comparison rather than a decomposition of total system persistence, which also depends on the structure of the network and the propagation of shocks through it, but it shows that the network component is large relative to own-firm persistence.

Columns (6) and (7) isolate the role of coefficient heterogeneity within the time-varying network specification. S--2SIV in column (6) uses the same annual
product-market weighting matrices as S--MGIV in column (7) but imposes common coefficients across firms. The estimated own-lag coefficient is 0.229 in column (6), compared with 0.346 in column (7), while the network coefficient is 0.328 rather than 0.283. Imposing common coefficients therefore shifts the estimated dynamic dependence away from firms' own leverage histories and towards the network. The difference is particularly pronounced for own-firm persistence, indicating that allowing firms to differ in their adjustment dynamics matters for how the model attributes leverage persistence between firms' own financing histories and the financing choices of connected firms.

Columns (5) and (7) show how the S--MGIV estimates change when the product-market network is held fixed rather than allowed to vary annually. The own-lag coefficient is 0.331 under the fixed network in column (5), close to the estimate of 0.346 under the time-varying network in column (7). The network coefficient, however, falls from 0.570 to 0.283, a reduction of approximately one-half. Allowing product-market relationships to evolve therefore has little effect on estimated own-firm persistence but materially changes the estimated strength of network dependence. This finding is consistent with the descriptive evidence in Table~\ref{tab:network_variation}: product-market connections are persistent but not fixed. It also relates to \citet{BreitungMuller2025}, who emphasise the
importance of allowing economically relevant firm networks to evolve as firms' business activities and competitive positions change. In our setting, averaging
these connections over the sample produces a  substantially stronger estimated network response than weighting past financing choices using the product-market
relationships prevailing at the time.

Finally, columns (1)--(3) show how estimated own-firm persistence changes as endogeneity and coefficient heterogeneity are addressed in models without network effects. The coefficient on lagged leverage is 0.785 under FE in column (1), falls to 0.508 under 2SIV in column (2), and falls further to 0.462 under MGIV in column (3). The implied adjustment speeds are 21.5\%,
49.2\%, and 53.8\%, respectively. Thus, even before network dependence is introduced, accounting for endogeneity and allowing leverage dynamics to differ across firms substantially reduces the persistence attributed to firms' own leverage histories.

\subsection{Direct, indirect and total marginal effects}

Table~\ref{tab:marginal_effects} reports the direct, indirect and total marginal effects implied by the network specifications. The direct effect measures the average response of a firm's own leverage to a change in a covariate, accounting for the transmission generated by the model. The indirect effect measures the associated response of leverage elsewhere in the product-market network, while the total effect combines the two. This decomposition is important because the coefficients on own and network-weighted characteristics do not by themselves measure how a change in a firm's fundamentals is transmitted through its economic connections.

We begin with S--MGIV under the time-varying network in specification (3). The direct, indirect and total effects of cash are $-0.164$, $-0.124$ and $-0.288$, respectively (all p$<$0.01). Thus, approximately 43\% of the total effect is associated with the network component. Tangibility also has
positive and statistically significant direct and indirect effects, at 0.198 and 0.150, respectively, producing a total effect of 0.349 ($p<0.01$). Profitability exhibits a particularly strong network component. Its direct
effect is $-0.657$, while its indirect effect is $-0.674$, yielding a total effect of $-1.332$ (all $p<0.01$). The indirect effect therefore accounts for approximately 51\% of the total response and is slightly larger in absolute magnitude than the direct effect. Market-to-book displays a different pattern. Its direct effect is positive and statistically significant at 0.005, whereas its indirect effect is negative at $-0.005$ and statistically insignificant. The total effect remains positive and statistically significant at 0.008. Unlike profitability, therefore, there is no evidence that the effect associated with market-to-book is reinforced through the product-market network.

The signs of the marginal effects for cash, profitability and tangibility are consistent with familiar capital-structure mechanisms. Greater cash holdings are associated with lower leverage, consistent with the availability of internal liquidity, while greater profitability is associated with lower leverage, consistent with greater reliance on internally generated funds. The positive effect of tangibility accords with the role of collateral in supporting debt capacity. In each case, the direct and indirect effects have the same sign, so the response elsewhere in the product-market network reinforces the own-firm response. Firm size behaves differently. Under the time-varying network in specification (3), its direct effect is positive and statistically significant at 0.084, while the indirect effect is economically small and statistically insignificant at 0.004. The resulting total effect, 0.087, is therefore almost entirely attributable to the direct component. Network transmission is consequently not a mechanical feature of every leverage determinant. It is economically important for cash, profitability and tangibility, but there is little evidence of a corresponding network component for firm size.

We next compare the time-varying S--MGIV results in specification (3) with the corresponding fixed-network results in specification (1). The contrast is substantial. Under S--MGIV with a fixed network, the direct effects of cash, profitability and tangibility are $-0.173$, $-0.660$ and 0.164, respectively, broadly similar to their counterparts under the time-varying network. The corresponding indirect effects, however, are $-0.978$, $-2.069$ and 0.932, compared with $-0.124$, $-0.674$ and 0.150 in specification (3). In absolute value, the indirect effects are approximately 5.7, 3.1 and 5.7 times their respective direct effects under the fixed network, compared with approximately 0.76, 1.03 and 0.76 under the time-varying network. Holding product-market connections fixed therefore has relatively little effect on the estimated direct responses but generates substantially stronger transmission through the network.

The contrast between specifications (1) and (3) is particularly striking for firm size. Under the fixed network, a positive direct effect of 0.090 (p$<$0.01) is accompanied by a large negative indirect effect of $-0.606$ (p$<$0.01), producing a negative total effect of $-0.516$ (p$<$0.01). With annual product-market weights, the direct effect remains positive at 0.084 (p$<$0.01), but the indirect effect falls to 0.004 and is statistically insignificant, leaving a positive total effect of 0.087 (p$<$0.01). Thus, holding product-market connections fixed not only increases the magnitude of network transmission but, for size, changes the sign of the estimated total effect.

These results also speak to the amplification mechanism emphasised in the peer-effects literature. \citet{LearyRoberts2014} show that interactions among firms can amplify the effects of firm-level determinants on corporate leverage. Our marginal-effects decomposition shows that such amplification can be economically important, but also that its magnitude is highly sensitive to how the product-market network is represented. Consistent with the importance of network specification emphasised by \citet{GrieserEtAl2022} and of allowing firm networks to evolve over time in \citet{BreitungMuller2025}, holding product-market connections fixed generates much larger indirect effects than using the relationships prevailing in each year.

Finally, the time-varying S--2SIV results in specification (2) provide the corresponding homogeneous-coefficient benchmark. The direct and indirect effects generally have the same signs as under S--MGIV in specification (3), although their magnitudes differ, consistent with the importance of coefficient heterogeneity documented in Table~\ref{tab:baseline_network}.

\subsection{Heterogeneity in network propagation across the firm-size distribution}
\label{sec:heterogeneity}

We next examine how the marginal effects under S--MGIV with the time-varying network vary across the firm-size distribution. Firms are assigned to quartiles according to size, with Q1 containing the smallest firms and Q4 the largest. Table~\ref{tab:size_heterogeneity} decomposes each full-sample marginal effect  into the contributions associated with firms in each size quartile. By construction, the quartile-specific contributions sum to the corresponding full-sample effect.

For cash holdings, the direct contribution is approximately constant across quartiles, at about $-0.041$, whereas the indirect contribution increases in absolute magnitude from $-0.0245$ in Q1 to $-0.0361$ in Q4. Consequently, the network share of the total contribution rises from 37.4\% among the smallest firms to 47.0\% among the largest, while the indirect-to-direct ratio increases from 0.60 to 0.89. Tangibility displays the same proportional gradient and indirect-to-direct ratio.\footnote{This equality follows from the specification rather than representing an independent empirical pattern. Since neither cash nor tangibility enters the model as a network-weighted covariate, their direct and indirect effects differ in scale but are transmitted through the same network structure.} 

Profitability exhibits a somewhat different pattern because both own and network-weighted profitability enter the leverage equation. Its direct contribution remains close to $-0.164$ across quartiles, whereas the indirect contribution increases in absolute magnitude from $-0.1333$ in Q1 to $-0.1964$ in Q4. The network share consequently rises from 44.7\% to 54.5\%, and the indirect-to-direct ratio from 0.81 to 1.20. For the largest firms, the network component therefore exceeds the own-firm component of the profitability effect.

Firm size provides a useful contrast, as its marginal-effect decomposition shows very little network transmission. The direct contribution accounts for almost all of the total effect in every quartile, while the network share rises only from 3.3\% in Q1 to 4.8\% in Q4 and the indirect-to-direct ratio remains between 0.03 and 0.05. Thus, the presence of a network-weighted covariate does not itself generate a large indirect effect.

In a nutshell, network propagation is therefore stronger among larger firms for several key leverage determinants, but is not a
uniform feature of the marginal effects. It is particularly pronounced for profitability and is also evident for cash and tangibility. By contrast, the network component of the firm-size effect remains negligible throughout the distribution.

\subsection{Spatial clustering}

An additional dimension of heterogeneity concerns whether firm-specific coefficients are systematically related to product-market proximity. We construct a measure of long-run product similarity by averaging the annual pairwise product-similarity measures over the sample period. Let $\bar{s}_{i,j}$ denote the resulting average similarity between firms $i$ and $j$. Before averaging, we set $s_{i,i,t}=1$, reflecting perfect own-firm product similarity. The off-diagonal elements of $\bar{\mathbf S}$ are strongly right-skewed, with a mean of 0.0138, a standard deviation of 0.0230, and a 97.5th percentile of 0.0741.

We transform product similarity into the standardized economic distance
\[
d_{i,j}
=
\frac{\max\{s^{*}-\bar{s}_{i,j},0\}}{\sigma_s},
\qquad i\neq j,
\]
where $s^{*}$ and $\sigma_s$ denote, respectively, the 97.5th percentile and standard deviation of the off-diagonal elements of $\bar{\mathbf S}$, and set $d_{ii}=0$. Product-market proximity is then measured using the Bartlett kernel
\[
k_{i,j}
=
\left(1-\frac{d_{i,j}}{h}\right)
\mathbbm{1}\{d_{i,j}<h\},
\qquad i\neq j,
\]
with $k_{i,i}=0$. Our baseline sets $h=1$, assigning positive weight to pairs with average product similarity above approximately 0.0511. The resulting matrix retains 5.3\% of all possible firm pairs, with a median of 22 connected firms per firm, and is row-normalized before calculating Moran's $I$.

For each firm-specific coefficient, we use Moran's $I$ to test whether firms that are closer in product-market space have more similar estimated responses. Statistical significance is assessed by permutation, randomly reallocating the coefficient estimates across firms while holding the proximity matrix fixed. Under random allocation, the expected value of Moran's $I$ is
$-1/(N-1)=-0.0018$.

Table~\ref{tab:moran_coefficients} provides evidence of product-market clustering in several economically important coefficients. The coefficient on own lagged leverage has $I=0.0427$ ($p=0.0136$), indicating that firms close in product-market space tend to exhibit more similar degrees of own-firm leverage persistence. The coefficients on tangibility, profitability, and
firm size also exhibit significant positive clustering, with Moran's $I$ of 0.0665, 0.0395, and 0.0770, respectively. By contrast, there is no statistically significant clustering in the coefficient on network-weighted lagged leverage ($I=0.0200$, $p=0.2128$).

These results show that firm-specific responses are not randomly distributed in product-market space. Similar firms exhibit similar responses along several dimensions, including own leverage dynamics and the effects of tangibility, profitability, and size, although this pattern does not extend to the response to connected firms' past leverage. This evidence supports allowing coefficients
to vary across firms: imposing common responses would suppress economically relevant heterogeneity that is itself systematically related to product-market proximity.

\section{Robustness checks}

\subsection{Book versus market leverage}

Book and market leverage capture different dimensions of firms' capital structures because market leverage incorporates variation in equity values that need not correspond to active financing decisions \citep{DeAngeloRoll2015}. We therefore examine whether the results depend on the definition of leverage. We therefore examine whether the results depend on the definition of leverage. Table~\ref{tab:robust_bl_ml} reports estimates from S--MGIV with the time-varying network using book and market leverage as alternative dependent variables.

The dynamic dependence on both own and connected firms’ past leverage remains economically and statistically important when leverage is measured at market rather than book values. The coefficient on network-weighted lagged leverage is positive and significant at the 1\% level in both specifications, declining from 0.283 for book leverage to 0.218 for market leverage. The coefficient on own lagged leverage similarly falls from 0.346 to 0.280. Thus, both own-firm persistence and dependence on the past leverage of economically connected firms are somewhat weaker when leverage is measured at market values. The lower own-lag coefficient is consistent with market leverage responding directly to changes in equity valuations even in the absence of active balance-sheet adjustment.

Importantly, the relative magnitude of the two dynamic coefficients is robust to the definition of leverage. Network-weighted lagged leverage represents approximately 45\% of the sum of the two mean dynamic coefficients for book leverage and 44\% for market leverage. The decline in the network coefficient under market leverage is therefore accompanied by a similar decline in own-firm persistence, preserving the relative importance of the two sources of dynamic dependence.

The firm-level controls are also broadly stable across the two leverage definitions, with one notable exception. Cash and profitability remain negative and highly significant, while tangibility and firm size remain positive and significant. Market-to-book, however, changes sign: its coefficient is 0.003 (p$<$0.01) for book leverage and $-0.016$ (p$<$0.01) for market leverage. The negative coefficient under market leverage is
consistent with the mechanical effect of higher equity valuations on the denominator of the market leverage ratio. The dividend indicator also becomes positive and significant under market leverage, whereas it is insignificant for book leverage.

The network-weighted controls display similar stability. Network-weighted firm size remains negative and significant at the 1\% level, with coefficients of $-0.022$ and $-0.015$ for book and market leverage, respectively. Network-weighted profitability remains insignificant under both definitions, while the weakly negative effect of network-weighted market-to-book under book leverage becomes statistically insignificant under market leverage.

\subsection{Network dependence under alternative specifications} \label{subsec:alternative_W}

The estimated strength of network dependence may depend on how the product-market network is constructed. Network estimates can be sensitive to the definition of economically relevant connections \citep{LearyRoberts2014,HobergPhillips2016JPE,GrieserEtAl2022}, while the connections themselves may evolve over time \citep{BreitungMuller2025}. Table~\ref{tab:alternative_spatial_specs} considers several
alternative definitions of the network, as well as additional industry controls and a specification that removes the connections generating the largest spillover effects.

The coefficient on network-weighted lagged leverage remains positive and statistically significant in every specification. Under S--MGIV with the time-varying product-market network, the estimate is 0.283. Adding industry fixed effects reduces the estimate to 0.234, although the difference from the baseline estimate is not statistically significant. We then consider two simpler network structures. The global linear-in-means specification treats all firms as neighbours and assigns each firm equal weight, $1/(N-1)$, whereas the industry linear-in-means specification connects firms within the same industry and
assigns equal weight to those connections. The estimated network coefficients under these two specifications are 0.413 and 0.380, respectively, and remain highly statistically significant.

The magnitudes of these coefficients are not directly comparable  across network definitions, since changing the weighting matrix changes the network-weighted leverage variable itself. The relevant finding is that dependence on connected firms' past leverage remains positive and statistically significant under substantially different definitions of economic connectivity.

A more demanding robustness exercise asks whether the baseline result is driven by the connections that generate the largest spillovers. We rank connections according to their estimated indirect marginal effects and remove the top 20\% before re-estimating the model. The coefficient on network-weighted lagged leverage falls from 0.283 to 0.164 but remains positive and significant at the 1\% level. Network dependence is therefore not
attributable solely to the connections generating the largest indirect effects; it remains detectable after these connections are excluded.

Own-firm persistence is also stable across these specifications. The coefficient on lagged leverage ranges from 0.291 to 0.346, compared with a much wider range of 0.164 to 0.413 for network-weighted lagged leverage. The exercises therefore support two features of the baseline results: own-firm persistence is preserved across alternative network constructions, and the positive role of connected firms' past leverage remains statistically significant under markedly different definitions of the network and after removing the largest spillover connections.

\subsection{Industry heterogeneity in network propagation}
\label{subsec:industry_heterogeneity}

Finally, we explore whether the strength of own-firm persistence and network dependence differs systematically across industries. Table~\ref{tab:industry_heterogeneity_full}
reports industry-specific Mean Group estimates for book leverage using the time-varying product-market network.

Own-firm persistence is positive and statistically significant in every industry. The coefficient on lagged leverage ranges from 0.269 in Retail to 0.406 in Mining, with estimates of similar order in Transportation, Finance/Insurance/Real Estate, Services, and Manufacturing. Gradual leverage adjustment therefore appears across all reported sectors rather than being concentrated in a small subset of industries.

The coefficient on network-weighted lagged leverage varies much more strongly across industries. It is largest in Services, at 0.701, and is also sizeable in Mining, Manufacturing, and Finance/Insurance/Real Estate, with estimates of 0.508, 0.309, and 0.225, respectively. By contrast, the estimates for Transportation and Retail are statistically insignificant, with the Retail estimate negative at $-0.384$. These results indicate substantial cross-industry heterogeneity in the strength of network dependence. 

The firm-level coefficients also vary across sectors.  Profitability is negative in all industries and statistically significant in most of them, while the effects of tangibility, capital expenditure, and firm size differ more substantially across groups. This cross-industry variation is consistent with
the broader evidence in the paper that firms differ materially in their leverage dynamics and responses to fundamentals.

\section{Concluding Remarks}

This paper examines whether corporate leverage persistence reflects only firms' own adjustment frictions or also the past financing choices of economically connected firms. We combine annual firm fundamentals with time-varying text-based product-market similarity data and embed the resulting network in a heterogeneous dynamic partial-adjustment model. The framework
allows own lagged leverage and network-weighted lagged leverage to enter jointly, while accommodating firm-specific responses and heterogeneous exposure to latent common shocks.

The results show that dependence on connected firms' past leverage is an economically important component of leverage dynamics. Under the time-varying network, the mean coefficient on network-weighted lagged leverage is sizeable relative to that on own lagged leverage, representing approximately 45\% of the sum of the two mean dynamic coefficients. When network dependence is omitted, substantially more persistence is attributed to the firm's own leverage history: introducing the network raises the implied mean speed of adjustment from 53.8\% to 65.4\%. Thus, part of the persistence conventionally attributed to firms' own adjustment frictions is associated with the past financing choices of economically connected firms.

A second implication concerns the identification of network effects. Product-market proximity is associated not only with firms' economic connections but also with similarities in their financing behaviour. The firm-specific estimates show significant product-market clustering in own-leverage persistence and in responses to several leverage determinants. To our knowledge, this provides the first evidence that firm-specific capital-structure responses themselves cluster in product-market space. Allowing for heterogeneous coefficients is therefore  economically relevant rather than simply an econometric generalisation: imposing common responses can suppress systematic differences in financing behaviour that are themselves related to the network. At the same time, the response to connected firms' past leverage does not exhibit statistically significant spatial clustering, distinguishing heterogeneity in the network response from the broader clustering of firm behaviour.

The results also show why the timing and structure of economic connections matter for measuring network dependence. Product-market relationships are persistent but evolve over time,and holding them fixed produces a substantially larger estimated network coefficient and much stronger indirect effects than using the relationships prevailing in each year. The relevant network for dynamic leverage adjustment is therefore not simply a long-run map of which firms are similar, but the set of economic relationships prevailing when the financing choices that subsequently propagate through the network are made.

The broader implication is that leverage persistence cannot be interpreted solely as a firm-level adjustment phenomenon. A firm's observed leverage reflects its own adjustment towards target leverage, the past financing choices transmitted through its product-market connections, systematic heterogeneity in financing behaviour, and exposure to common economic shocks. Separating these components requires a framework that accommodates evolving network relationships and allows firm-specific responses themselves to be spatially dependent. Doing so changes both the interpretation of conventional adjustment-speed estimates and the empirical requirements for identifying network effects in corporate finance.

\bibliographystyle{apalike}
\bibliography{references.bib}

\newpage

\begin{table}[htbp]
  \centering

  \caption{Summary Statistics}
  \label{tab:summary_stats}

  \renewcommand{\arraystretch}{1.05}
  \scalebox{0.90}{
  \begin{tabular}{llrrrr}
    \toprule
    Variable & Variation & Mean & Std. Dev. & Min & Max \\
    \midrule

    \multirow{3}{*}{Book leverage}
      & Overall & 0.234 & 0.180 & 0.000 & 0.804 \\
      & Within  &       & 0.115 &       &       \\
      & Between &       & 0.138 &       &       \\
    \midrule

    \multirow{3}{*}{Market leverage}
      & Overall & 0.168 & 0.148 & 0.000 & 0.649 \\
      & Within  &       & 0.097 &       &       \\
      & Between &       & 0.113 &       &       \\
    \midrule

    \multirow{3}{*}{Cash}
      & Overall & 0.127 & 0.162 & 0.000 & 0.848 \\
      & Within  &       & 0.096 &       &       \\
      & Between &       & 0.131 &       &       \\
    \midrule

    \multirow{3}{*}{Market-to-book}
      & Overall & 3.671 & 7.278 & -5.517 & 60.446 \\
      & Within  &       & 4.956 &        &        \\
      & Between &       & 5.334 &        &        \\
    \midrule

    \multirow{3}{*}{Tangibility}
      & Overall & 0.287 & 0.253 & 0.000 & 0.885 \\
      & Within  &       & 0.086 &       &       \\
      & Between &       & 0.238 &       &       \\
    \midrule

    \multirow{3}{*}{Profitability}
      & Overall & 0.136 & 0.158 & -0.232 & 1.287 \\
      & Within  &       & 0.077 &        &       \\
      & Between &       & 0.138 &        &       \\
    \midrule

    \multirow{3}{*}{SG\&A}
      & Overall & 0.179 & 0.157 & 0.000 & 0.737 \\
      & Within  &       & 0.075 &       &       \\
      & Between &       & 0.139 &       &       \\
    \midrule

    \multirow{3}{*}{CapEx}
      & Overall & 0.047 & 0.049 & 0.000 & 0.259 \\
      & Within  &       & 0.032 &       &       \\
      & Between &       & 0.038 &       &       \\
    \midrule

    \multirow{3}{*}{Size}
      & Overall & 7.280 & 2.478 & 1.665 & 12.529 \\
      & Within  &       & 0.923 &       &        \\
      & Between &       & 2.302 &       &        \\
    \midrule

    \multirow{3}{*}{Dividend dummy}
      & Overall & 0.718 & 0.450 & 0.000 & 1.000 \\
      & Within  &       & 0.259 &       &       \\
      & Between &       & 0.369 &       &       \\

    \bottomrule
  \end{tabular}
  }

  \vspace{0.4em}
    \begin{minipage}{0.95\textwidth}
    \footnotesize
    \textit{\textbf{Notes}:}
    This table reports descriptive statistics for the variables used in the empirical analysis. Book leverage is defined as total debt divided by total assets, $(\mathrm{DLTT}+\mathrm{DLC})/\mathrm{AT}$, where $\mathrm{DLTT}$ and $\mathrm{DLC}$ denote long-term debt and debt in current liabilities, respectively. Market leverage is defined as total debt divided by the market value of assets, $(\mathrm{DLTT}+\mathrm{DLC})/[(\mathrm{AT}-\mathrm{CEQ})+(\mathrm{PRCC\_F}\times\mathrm{CSHO})]$, where $\mathrm{CEQ}$ is common equity, $\mathrm{PRCC\_F}$ is the fiscal year-end share price, and $\mathrm{CSHO}$ is common shares outstanding. Cash is defined as cash and short-term investments divided by total assets ($\mathrm{CHE}/\mathrm{AT}$); Market-to-book ratio is defined as the market value of assets divided by total book assets, $\left[(\mathrm{AT}-\mathrm{CEQ})+ (\mathrm{PRCC\_F}\times\mathrm{CSHO})\right]/\mathrm{AT}$; Tangibility is net property, plant and equipment over total assets ($\mathrm{PPENT}/\mathrm{AT}$). Profitability is operating income before depreciation over total assets ($\mathrm{OIBDP}/\mathrm{AT}$). SG\&A is selling, general and administrative expenses over sales ($\mathrm{XSGA}/\mathrm{SALE}$). CapEx is capital expenditures divided by total assets ($\mathrm{CAPX}/\mathrm{AT}$). Size is the natural logarithm of total assets, $\ln(\mathrm{AT})$. Dividend dummy equals one when the firm reports positive common dividends ($\mathrm{DVC}>0$), and zero otherwise. Standard deviations are decomposed into overall, within-firm, and between-firm variation; means, minima, and maxima refer to the overall distribution. The balanced panel contains 549 firms and 18,666 firm-year observations over 1990--2023.
    \end{minipage}
\end{table}

\begin{table}[htbp]
\centering
\caption{Time Variation in Product-Market Connections}
\label{tab:network_variation}

\begin{tabular}{lclc}
\toprule
\multicolumn{2}{c}{Pairwise time-series variation}
&
\multicolumn{2}{c}{Year-to-year network variation}
\\
\cmidrule(lr){1-2} \cmidrule(lr){3-4}

Statistic & Value & Statistic & Value \\
\midrule

Average similarity
& 0.014
& Average correlation, $W_t$ and $W_{t-1}$
& 0.891
\\

Average within-pair SD
& 0.012
& Mean absolute annual change
& 0.006
\\

Median within-pair range
& 0.037
& Average annual RMSE
& 0.013
\\

Median coefficient of variation
& 1.617
& Share with material annual change
& 0.149
\\

\bottomrule
\end{tabular}

\vspace{0.5em}

\begin{minipage}{0.95\textwidth}
\footnotesize
\textit{\textbf{Notes}:}
The left panel reports time-series variation in product similarity across
150,426 unique firm pairs over 1990--2023. For each pair $(i,j)$, the
within-pair standard deviation, range, and coefficient of variation are
calculated from its annual product-similarity measures. The coefficient of
variation is the time-series standard deviation divided by the time-series
mean and is calculated for pairs with non-zero mean similarity. The right
panel reports changes in the product-market network between consecutive
years. The annual network correlation is the correlation between the
off-diagonal elements of the similarity matrices in years $t$ and $t-1$.
A material annual change is defined as an absolute change in pairwise
similarity exceeding one-half of the overall standard deviation of annual
off-diagonal similarities.
\end{minipage}
\end{table}

\begin{table}[htbp]
\centering
\caption{Partial adjustment with spatial and time-varying network effects}
\label{tab:baseline_network}
\resizebox{\textwidth}{!}{%
\begin{tabular}{lccccccc}
\hline
\hline
& \multicolumn{3}{c}{No network effects}
& \multicolumn{2}{c}{Fixed network}
& \multicolumn{2}{c}{Time-varying network} \\
\cline{2-4} \cline{5-6} \cline{7-8}
& FE & 2SIV & MGIV
& S--2SIV & S--MGIV
& S--2SIV & S--MGIV \\
& (1) & (2) & (3) & (4) & (5) & (6) & (7) \\
\hline

$L.\mathit{Leverage}$ 
 & 0.785*** & 0.508*** & 0.462*** & 0.257*** & 0.331*** & 0.229*** & 0.346*** \\
 & (0.005)  & (0.089)  & (0.015)  & (0.039)  & (0.013)  & (0.041)  & (0.013)  \\
[0.4em]

$\mathit{Cash}$ 
 & -0.063*** & -0.077** & -0.102*** & -0.148*** & -0.114*** & -0.147*** & -0.107*** \\
 & (0.006)   & (0.037)  & (0.034)   & (0.015)   & (0.022)   & (0.015)   & (0.021)   \\
[0.4em]

$\mathit{Market\mbox{-}to\mbox{-}book}$ 
 & 0.001*** & 0.002** & 0.004*** & 0.001*** & 0.004*** & 0.001*** & 0.003*** \\
 & (0.000)  & (0.001) & (0.001)  & (0.000)  & (0.001)  & (0.000)  & (0.001)  \\
[0.4em]

$\mathit{Tangibility}$ 
 & 0.034*** & 0.049** & 0.153*** & 0.115*** & 0.109** & 0.112*** & 0.130** \\
 & (0.006)  & (0.021) & (0.051) & (0.028)  & (0.049) & (0.027)  & (0.050)  \\
[0.4em]

$\mathit{Profitability}$ 
 & -0.152*** & -0.164*** & -0.376*** & -0.179*** & -0.438*** & -0.182*** & -0.429*** \\
 & (0.007)   & (0.043)   & (0.043)   & (0.023)   & (0.038)   & (0.024)   & (0.035)   \\
[0.4em]

$\mathit{SG\&A}$ 
 & -0.007 & 0.027 & 0.038 & 0.006 & -0.001 & -0.026 & -0.038 \\
 & (0.007) & (0.028) & (0.078) & (0.023) & (0.066) & (0.025) & (0.062) \\
[0.4em]

$\mathit{CapEx}$ 
 & 0.081*** & 0.111 & -0.109 & -0.093** & 0.002 & -0.146*** & -0.027 \\
 & (0.017)  & (0.093) & (0.101) & (0.036)  & (0.077) & (0.040)   & (0.081) \\
[0.4em]

$\mathit{Size}$ 
 & 0.010*** & 0.017*** & 0.038*** & 0.031*** & 0.061*** & 0.029*** & 0.055*** \\
 & (0.001)  & (0.005)  & (0.006)  & (0.005)  & (0.006)  & (0.005)  & (0.005)  \\
[0.4em]

$\mathit{Dividend\ dummy}$ 
 & 0.008*** & 0.005 & -0.022 & -0.024*** & 0.029 & -0.028*** & 0.029 \\
 & (0.002)  & (0.008) & (0.042) & (0.005)   & (0.036) & (0.005)   & (0.035) \\

\hline

$W L.\mathit{Leverage}$ 
 &  &  &  & 0.748*** & 0.570*** & 0.328*** & 0.283*** \\
 &  &  &  & (0.153)  & (0.133)  & (0.101)  & (0.072)  \\
[0.4em]

$W\,\mathit{Market\mbox{-}to\mbox{-}book}$ 
 &  &  &  & -0.002 & -0.007** & 0.003* & -0.003* \\
 &  &  &  & (0.002) & (0.003) & (0.002) & (0.002) \\
[0.4em]

$W\,\mathit{Size}$ 
 &  &  &  & -0.062** & -0.112*** & -0.007 & -0.022*** \\
 &  &  &  & (0.029) & (0.038) & (0.005) & (0.006) \\
[0.4em]

$W\,\mathit{Profitability}$ 
 &  &  &  & -0.249** & 0.168 & -0.090 & -0.066 \\
 &  &  &  & (0.097) & (0.144) & (0.083) & (0.082) \\

\hline
$N_g$ & 549 & 549 & 549 & 549 & 549 & 549 & 549 \\
$N$   & 18117 & 17568 & 17568 & 18117 & 18117 & 18117 & 17568 \\
$\chi^2_J$ &  & 11.586 &  & 47.142 &  & 42.964 &  \\
$p_J$      &  & 0.115  &  & 0.000  &  & 0.000  &  \\
factors in X      &  & 1  & 1 & 1 & 1 & 1 & 1 \\
factors in y      &  & 1  &  & 1 &  & 1 &  \\
\hline
\hline
IV estimation
& No & Yes & Yes & Yes & Yes & Yes & Yes \\
Network effects
& No & No & No & Yes & Yes & Yes & Yes \\
Time-varying $W_t$
& No & No & No & No & No & Yes & Yes \\
Coefficient heterogeneity
& No & No & Yes & No & Yes & No & Yes \\
\hline
\end{tabular}%
}
\vspace{0.5em}

\begin{minipage}{1\textwidth}
\footnotesize
\textit{\textbf{Notes}:}
The dependent variable is book leverage. FE denotes estimation with firm and year fixed effects. 2SIV and MGIV denote two-stage IV and mean-group IV estimators, respectively. S--2SIV and S--MGIV denote their spatial counterparts, which augment the model with network-weighted lagged leverage and network-weighted covariates. Columns under ``Fixed network'' use a
time-invariant product-market weighting matrix $W$, whereas columns under ``Time-varying network'' use annual weighting matrices $W_t$. $L.\mathit{Leverage}$ captures own leverage persistence, while $W L.\mathit{Leverage}$ measures the effect of connected firms' lagged leverage on current leverage. $W\,\mathit{Market\mbox{-}to\mbox{-}book}$,
$W\,\mathit{Size}$, and $W\,\mathit{Profitability}$ capture contextual network effects associated with connected firms' characteristics. $N_g$ denotes the
number of firms and $N$ the number of firm-year  observations. $\chi^2_J$ and $p_J$ report the Hansen-type test of overidentifying restrictions where applicable. Standard errors are reported in parentheses.
* p$<$0.10, ** p$<$0.05, *** p$<$0.01.
\end{minipage}
\end{table}

\begin{landscape}
\begin{table}[!htbp]
\centering
\caption{Direct, indirect and total marginal effects of covariates}
\label{tab:marginal_effects}
\small
\setlength{\tabcolsep}{4.5pt}
\renewcommand{\arraystretch}{1.05}

\begin{tabular}{lccccccccc}
\hline\hline
&
\multicolumn{3}{c}{Fixed network}
&
\multicolumn{6}{c}{Time-varying network}
\\
\cline{2-4}\cline{5-10}
&
\multicolumn{3}{c}{S--MGIV}
&
\multicolumn{3}{c}{S--2SIV}
&
\multicolumn{3}{c}{S--MGIV}
\\
&
\multicolumn{3}{c}{(1)}
&
\multicolumn{3}{c}{(2)}
&
\multicolumn{3}{c}{(3)}
\\
\cline{2-4}\cline{5-7}\cline{8-10}
&
Direct & Indirect & Total
&
Direct & Indirect & Total
&
Direct & Indirect & Total
\\
\hline

$\mathit{Cash}$
& $-0.173^{***}$ & $-0.978^{***}$ & $-1.151^{***}$
& $-0.191^{***}$ & $-0.141^{***}$ & $-0.332^{***}$
& $-0.164^{***}$ & $-0.124^{***}$ & $-0.288^{***}$ \\
& $(0.034)$ & $(0.034)$ & $(0.034)$
& $(0.033)$ & $(0.038)$ & $(0.134)$
& $(0.039)$ & $(0.023)$ & $(0.113)$ \\[2pt]

$\mathit{Market\mbox{-}to\mbox{-}book}$
& $0.006$ & $-0.028$ & $-0.022$
& $0.001$ & $0.008$ & $0.009$
& $0.005^{***}$ & $-0.005$ & $0.008^{***}$ \\
& $(0.034)$ & $(0.034)$ & $(0.034)$
& $(0.001)$ & $(0.007)$ & $(0.012)$
& $(0.001)$ & $(0.004)$ & $(0.003)$ \\[2pt]

$\mathit{Tangibility}$
& $0.164^{**}$ & $0.932^{***}$ & $1.096^{***}$
& $0.145$ & $0.107^{***}$ & $0.252$
& $0.198^{***}$ & $0.150^{***}$ & $0.349^{***}$ \\
& $(0.073)$ & $(0.073)$ & $(0.073)$
& $(0.090)$ & $(0.041)$ & $(0.253)$
& $(0.079)$ & $(0.041)$ & $(0.118)$ \\[2pt]

$\mathit{Profitability}$
& $-0.660^{***}$ & $-2.069^{***}$ & $-2.729^{***}$
& $-0.237^{***}$ & $-0.376^{***}$ & $-0.613^{***}$
& $-0.657^{***}$ & $-0.674^{***}$ & $-1.332^{***}$ \\
& $(0.109)$ & $(0.109)$ & $(0.109)$
& $(0.073)$ & $(0.011)$ & $(0.238)$
& $(0.073)$ & $(0.152)$ & $(0.526)$ \\[2pt]

$\mathit{SG\&A}$
& $-0.001$ & $-0.007$ & $-0.008$
& $-0.034$ & $-0.025$ & $-0.059$
& $-0.059$ & $-0.045$ & $-0.103$ \\
& $(0.099)$ & $(0.099)$ & $(0.099)$
& $(0.086)$ & $(0.018)$ & $(0.104)$
& $(0.063)$ & $(0.049)$ & $(0.126)$ \\[2pt]

$\mathit{CapEx}$
& $0.004$ & $0.020$ & $0.024$
& $-0.190$ & $-0.140$ & $-0.330$
& $-0.041$ & $-0.031$ & $-0.072$ \\
& $(0.117)$ & $(0.117)$ & $(0.117)$
& $(0.241)$ & $(0.264)$ & $(0.290)$
& $(0.044)$ & $(0.126)$ & $(0.092)$ \\[2pt]

$\mathit{Size}$
& $0.090^{***}$ & $-0.606^{***}$ & $-0.516^{***}$
& $0.038^{***}$ & $0.012$ & $0.050$
& $0.084^{***}$ & $0.004$ & $0.087^{***}$ \\
& $(0.009)$ & $(0.009)$ & $(0.009)$
& $(0.009)$ & $(0.041)$ & $(0.073)$
& $(0.008)$ & $(0.005)$ & $(0.021)$ \\[2pt]

$\mathit{Dividend\ dummy}$
& $0.045$ & $0.253^{***}$ & $0.297^{***}$
& $-0.036$ & $-0.027$ & $-0.063$
& $0.044$ & $0.033$ & $0.077$ \\
& $(0.055)$ & $(0.055)$ & $(0.055)$
& $(0.043)$ & $(0.084)$ & $(0.141)$
& $(0.052)$ & $(0.031)$ & $(0.072)$ \\

\hline
$N_g$
& \multicolumn{3}{c}{549}
& \multicolumn{3}{c}{549}
& \multicolumn{3}{c}{549} \\

$N$
& \multicolumn{3}{c}{18117}
& \multicolumn{3}{c}{18117}
& \multicolumn{3}{c}{17568} \\
\hline\hline
\end{tabular}

\vspace{0.5em}

\begin{minipage}{0.98\linewidth}
\footnotesize
\textit{\textbf{Notes}:}
The table reports direct, indirect and total marginal effects on book leverage. S--MGIV under ``Fixed network'' uses a time-invariant product-market weighting
matrix, while S--2SIV and S--MGIV under ``Time-varying network'' use annual product-market weighting matrices. Direct effects measure the average response of the focal firm's leverage to a change in its own covariate. Indirect effects measure the associated response of leverage elsewhere in the product-market network, and total effects are the sum of the direct and indirect effects. $N_g$ denotes the number of firms and $N$ the number of firm-year observations. Standard errors are reported in parentheses. * p$<$0.10, ** p$<$0.05, *** p$<$0.01.
\end{minipage}

\end{table}
\end{landscape}

\begin{table}[htbp]
\centering
\caption{Direct and indirect marginal effects by firm size quartile}
\label{tab:size_heterogeneity}
\resizebox{0.92\textwidth}{!}{%
\begin{tabular}{lccccc}
\hline\hline
 & All firms & Q1 (smallest) & Q2 & Q3 & Q4 (largest) \\
\hline

\multicolumn{6}{l}{\textit{Panel A: Cash holdings}} \\[0.2em]
Total effect
    & -0.2876 & -0.0656 & -0.0704 & -0.0746 & -0.0770 \\
Direct effect
    & -0.1635 & -0.0411 & -0.0408 & -0.0408 & -0.0408 \\
Indirect effect
    & -0.1240 & -0.0245 & -0.0296 & -0.0338 & -0.0361 \\
Indirect / total (\%)
    & 43.1 & 37.4 & 42.1 & 45.3 & 47.0 \\
Indirect / direct
    & 0.76 & 0.60 & 0.73 & 0.83 & 0.89 \\
\hline

\multicolumn{6}{l}{\textit{Panel B: Tangibility}} \\[0.2em]
Total effect
    & 0.3488 & 0.0796 & 0.0854 & 0.0905 & 0.0933 \\
Direct effect
    & 0.1984 & 0.0499 & 0.0495 & 0.0495 & 0.0495 \\
Indirect effect
    & 0.1505 & 0.0297 & 0.0359 & 0.0410 & 0.0438 \\
Indirect / total (\%)
    & 43.0 & 37.4 & 42.1 & 45.3 & 47.0 \\
Indirect / direct
    & 0.76 & 0.60 & 0.73 & 0.83 & 0.89 \\
\hline

\multicolumn{6}{l}{\textit{Panel C: Profitability}} \\[0.2em]
Total effect
    & -1.3315 & -0.2985 & -0.3250 & -0.3476 & -0.3605 \\
Direct effect
    & -0.6573 & -0.1652 & -0.1640 & -0.1640 & -0.1641 \\
Indirect effect
    & -0.6743 & -0.1333 & -0.1610 & -0.1835 & -0.1964 \\
Indirect / total (\%)
    & 50.6 & 44.7 & 49.6 & 52.8 & 54.5 \\
Indirect / direct
    & 1.03 & 0.81 & 0.98 & 1.12 & 1.20 \\
\hline

\multicolumn{6}{l}{\textit{Panel D: Firm size}} \\[0.2em]
Total effect
    & 0.0875 & 0.0218 & 0.0218 & 0.0219 & 0.0220 \\
Direct effect
    & 0.0838 & 0.0211 & 0.0209 & 0.0209 & 0.0209 \\
Indirect effect
    & 0.0036 & 0.0007 & 0.0009 & 0.0010 & 0.0011 \\
Indirect / total (\%)
    & 4.6 & 3.3 & 4.0 & 4.5 & 4.8 \\
Indirect / direct
    & 0.05 & 0.03 & 0.04 & 0.05 & 0.05 \\
\hline\hline
\end{tabular}%
}

\vspace{0.5em}

\begin{minipage}{{1}\textwidth}
\footnotesize
\textit{Notes:} The table decomposes the full-sample marginal effects implied
by the preferred $S$--MGIV$_{W_t}$ specification across quartiles of the firm
size distribution. Q1 contains the smallest firms and Q4 the largest.
Quartile-specific direct, indirect, and total effects represent each
quartile's contribution to the corresponding aggregate marginal effect and,
by construction, sum to the full-sample effect. ``Indirect / total'' reports
the absolute magnitude of the indirect contribution as a percentage of the
absolute total contribution within each quartile, while ``Indirect / direct''
reports the ratio of the absolute indirect contribution to the absolute
direct contribution.
\end{minipage}
\end{table}

\begin{table}[!htbp]
\centering
\caption{Product-Market Clustering of Firm-Specific Coefficients}
\label{tab:moran_coefficients}
\begin{tabular}{lcc}
\hline\hline
Coefficient & Moran's $I$ & Permutation $p$-value \\
\hline
Own lagged leverage               & 0.0427 & 0.0136 \\
Network-weighted lagged leverage  & 0.0200 & 0.2128 \\
Tangibility                       & 0.0665 & 0.0023 \\
Profitability                     & 0.0395 & 0.0192 \\
Firm size                         & 0.0770 & 0.0002 \\
\hline\hline
\end{tabular}

\begin{minipage}{0.90\textwidth}
\footnotesize
\textit{Notes:} This table reports Moran's $I$ statistics for selected
firm-specific coefficient estimates. Product-market proximity is constructed
from the time-series average of annual pairwise product-similarity measures.
Economic distance is defined as
$d_{ij}=\max\{s^{*}-\bar{s}_{ij},0\}/\sigma_s$, where $s^{*}$ is the
97.5th percentile of the off-diagonal average similarity distribution and
$\sigma_s$ is its standard deviation. A Bartlett kernel with bandwidth
$h=1$ is used to construct the proximity matrix, which is subsequently
row-normalized. The resulting matrix retains 5.3\% of all possible
off-diagonal links and has a median of 22 connected firms per firm.
Permutation $p$-values are obtained by randomly reallocating the estimated
coefficients across firms while holding the product-market proximity matrix
fixed. Under the null of random allocation, the expected value of Moran's
$I$ is $-1/(N-1)=-0.0018$, with $N=549$. %Results for all firm-specific coefficients are reported in Table~\ref{tab:moran_full}.
\end{minipage}
\end{table}

\begin{table}[htbp]
\centering
\caption{Robustness: Book versus market leverage}
\label{tab:robust_bl_ml}

\resizebox{0.92\textwidth}{!}{%
\begin{tabular}{lcc}
\hline\hline
 & S--MGIV$_{W_t}$ (Book leverage: $bl$)
 & S--MGIV$_{W_t}$ (Market leverage: $ml$) \\
\hline

\multicolumn{3}{l}{\textit{Panel A: Own-firm dynamics and characteristics}} \\[0.3em]

$L.\mathit{Leverage}$
& 0.346*** & 0.280*** \\
& (0.013) & (0.012) \\[0.3em]

$\mathit{Cash}$
& -0.107*** & -0.061*** \\
& (0.021) & (0.019) \\

$\mathit{Market\mbox{-}to\mbox{-}book}$
& 0.003*** & -0.016*** \\
& (0.001) & (0.001) \\

$\mathit{Tangibility}$
& 0.130** & 0.109** \\
& (0.050) & (0.043) \\

$\mathit{Profitability}$
& -0.429*** & -0.392*** \\
& (0.035) & (0.033) \\

$\mathit{SG\&A}$
& -0.038 & -0.100* \\
& (0.062) & (0.058) \\

$\mathit{CapEx}$
& -0.027 & 0.025 \\
& (0.081) & (0.071) \\

$\mathit{Size}$
& 0.055*** & 0.056*** \\
& (0.005) & (0.004) \\

$\mathit{Dividend\ dummy}$
& 0.029 & 0.079*** \\
& (0.035) & (0.027) \\[0.4em]

\hline

\multicolumn{3}{l}{\textit{Panel B: Network effects}} \\[0.3em]

$W_t L.\mathit{Leverage}$
& 0.283*** & 0.218*** \\
& (0.072) & (0.070) \\

$W_t\,\mathit{Market\mbox{-}to\mbox{-}book}$
& -0.003* & -0.002 \\
& (0.002) & (0.002) \\

$W_t\,\mathit{Size}$
& -0.022*** & -0.015*** \\
& (0.006) & (0.005) \\

$W_t\,\mathit{Profitability}$
& -0.066 & -0.034 \\
& (0.082) & (0.078) \\

\hline

Number of firms
& 549 & 549 \\

Observations
& 17{,}568 & 17{,}568 \\

\hline\hline
\end{tabular}%
}

\vspace{0.5em}

\begin{minipage}{1\textwidth}
\footnotesize
\textit{Notes:} This table examines whether the main results are robust to the
definition of leverage. Both columns are estimated using the preferred
S--MGIV$_{W_t}$ specification. Column (1) uses book leverage and column (2)
uses market leverage as the dependent variable.
$L.\mathit{Leverage}$ denotes the lagged dependent variable, while
$W_tL.\mathit{Leverage}$ denotes the network-weighted lagged leverage of
product market peers constructed using the time-varying weighting matrix
$W_t$. Panel A reports own-firm dynamics and firm characteristics. Panel B
reports contextual network effects associated with peers' market-to-book
ratios, size and profitability. Both specifications use the same time-varying
product market network and the same set of controls. Standard errors are
reported in parentheses. *, **, and *** denote statistical significance at
the 10\%, 5\%, and 1\% levels, respectively.
\end{minipage}

\end{table}

\begin{table}[htbp]
\centering
\caption{Alternative specifications of the spatial partial-adjustment model}
\label{tab:alternative_spatial_specs}

\resizebox{\textwidth}{!}{%
\begin{tabular}{lccccc}
\hline\hline
& Baseline $W_t$
& $W_t$ + Industry
& Group linear
& Industry linear
& Excl. top 20\% peers \\
& (1) & (2) & (3) & (4) & (5) \\
\hline

\multicolumn{6}{l}{\textit{Panel A: Own-firm dynamics and characteristics}} \\[0.3em]

$L.\mathit{Book\ leverage}$
& 0.346***
& 0.344***
& 0.304***
& 0.291***
& 0.340*** \\
& (0.013)
& (0.013)
& (0.013)
& (0.013)
& (0.013) \\[0.2em]

$\mathit{Cash}$
& -0.107***
& -0.105***
& -0.123***
& -0.122***
& -0.101*** \\
& (0.021)
& (0.020)
& (0.028)
& (0.026)
& (0.022) \\

$\mathit{Market\mbox{-}to\mbox{-}book}$
& 0.003***
& 0.003***
& 0.003**
& 0.003**
& 0.003*** \\
& (0.001)
& (0.001)
& (0.001)
& (0.001)
& (0.001) \\

$\mathit{Tangibility}$
& 0.130**
& 0.081*
& -0.094
& -0.059
& 0.104** \\
& (0.050)
& (0.043)
& (0.110)
& (0.091)
& (0.049) \\

$\mathit{Profitability}$
& -0.429***
& -0.392***
& -0.384***
& -0.426***
& -0.434*** \\
& (0.035)
& (0.030)
& (0.044)
& (0.047)
& (0.036) \\

$\mathit{SG\&A}$
& -0.038
& -0.029
& -0.123**
& -0.107**
& -0.059 \\
& (0.062)
& (0.048)
& (0.056)
& (0.054)
& (0.063) \\

$\mathit{CapEx}$
& -0.027
& -0.003
& 0.274
& 0.260
& -0.031 \\
& (0.081)
& (0.099)
& (0.188)
& (0.176)
& (0.082) \\

$\mathit{Size}$
& 0.055***
& 0.057***
& 0.075***
& 0.076***
& 0.053*** \\
& (0.005)
& (0.005)
& (0.006)
& (0.006)
& (0.005) \\

$\mathit{Dividend\ dummy}$
& 0.029
& -0.018
& -0.000
& 0.000
& 0.033 \\
& (0.035)
& (0.029)
& (0.002)
& (0.003)
& (0.037) \\[0.4em]

\hline

\multicolumn{6}{l}{\textit{Panel B: Network effects}} \\[0.3em]

$W L.\mathit{Book\ leverage}$
& 0.283***
& 0.234***
& 0.413***
& 0.380***
& 0.164*** \\
& (0.072)
& (0.085)
& (0.064)
& (0.058)
& (0.052) \\

$W\,\mathit{Market\mbox{-}to\mbox{-}book}$
& -0.003*
& -0.001
& -0.001
& -0.002*
& -0.004* \\
& (0.002)
& (0.002)
& (0.001)
& (0.001)
& (0.002) \\

$W\,\mathit{Size}$
& -0.022***
& -0.015***
& -0.072***
& -0.075***
& -0.005 \\
& (0.006)
& (0.006)
& (0.008)
& (0.009)
& (0.005) \\

$W\,\mathit{Profitability}$
& -0.066
& -0.168*
& 0.335***
& 0.360***
& 0.011 \\
& (0.082)
& (0.088)
& (0.080)
& (0.076)
& (0.078) \\

\hline

Number of firms
& 549 & 549 & 549 & 549 & 549 \\

Observations
& 17,568 & 17,568 & 17,568 & 17,568 & 17,568 \\

\hline\hline
\end{tabular}%
}

\vspace{0.5em}

\begin{minipage}{1\textwidth}
\footnotesize
\textit{Notes:} This table reports S--MGIV estimates of the dynamic
partial-adjustment model under alternative network specifications. Column (1) uses the time-varying product-market weighting matrix. Column (2) additionally includes industry fixed effects. Columns (3) and (4) replace the product-market network with group-linear and industry-linear weighting structures, respectively. Column (5) reconstructs the time-varying product-market network after excluding the top 20\% of highest-weight connections from each firm's network. The firm-year sample is unchanged across columns. Panel A reports own-firm dynamics and characteristics, while Panel B reports the network-weighted lagged dependent variable and network-weighted firm characteristics. Standard errors are reported in parentheses. *, **, and *** denote statistical significance at the 10\%, 5\%, and 1\% levels, respectively.
\end{minipage}

\end{table}

\begin{table}[htbp]
\centering
\caption{Industry Heterogeneity in Leverage Dynamics and Network Effects}
\label{tab:industry_heterogeneity_full}

\resizebox{\textwidth}{!}{%
\begin{tabular}{lcccccc}
\toprule
 & Mining & Manufact. & Transport. & Retail 
 & Fin., Ins., Real Est. & Services \\
\midrule

$L.\mathit{Leverage}$
& 0.406*** & 0.309*** & 0.403*** & 0.269*** & 0.393*** & 0.330*** \\
& (0.057)  & (0.018)  & (0.035)  & (0.065)  & (0.033)  & (0.037)  \\

$\mathit{Cash}$
& -0.043   & -0.101*** & -0.127*  & -0.117   & -0.092   & -0.151** \\
& (0.111)  & (0.027)   & (0.071)  & (0.114)  & (0.065)  & (0.059)  \\

$\mathit{Market\mbox{-}to\mbox{-}book}$
& -0.006   & 0.006*** & 0.003    & 0.000    & -0.002   & 0.004* \\
& (0.005)  & (0.001)  & (0.002)  & (0.003)  & (0.002)  & (0.002) \\

$\mathit{Tangibility}$
& 0.071    & -0.019   & 0.086*   & 0.245**  & 0.641*** & 0.097 \\
& (0.109)  & (0.056)  & (0.050)  & (0.122)  & (0.227)  & (0.124) \\

$\mathit{Profitability}$
& -0.349*** & -0.487*** & -0.486*** & -0.491*** & -0.173   & -0.494*** \\
& (0.047)   & (0.039)   & (0.086)   & (0.174)   & (0.151)  & (0.088)   \\

$\mathit{SG\&A}$
& 0.248    & -0.025   & 0.134    & -0.199   & -0.250   & -0.200 \\
& (0.435)  & (0.072)  & (0.237)  & (0.236)  & (0.174)  & (0.148) \\

$\mathit{CapEx}$
& 0.180*   & -0.014   & 0.286*** & -0.007   & -0.376   & -0.315 \\
& (0.100)  & (0.097)  & (0.096)  & (0.151)  & (0.358)  & (0.220) \\

$\mathit{Size}$
& 0.030    & 0.071*** & 0.035*** & 0.053    & 0.035*** & 0.043** \\
& (0.018)  & (0.007)  & (0.012)  & (0.032)  & (0.011)  & (0.018) \\

$\mathit{Dividend\ dummy}$
& 0.025    & -0.089   & 0.185*** & 0.131    & 0.194**  & 0.082 \\
& (0.161)  & (0.057)  & (0.056)  & (0.188)  & (0.089)  & (0.086) \\

\midrule

$W_{t-1}L.\mathit{Leverage}$
& 0.508    & 0.309**  & 0.096    & -0.384   & 0.225**  & 0.701*** \\
& (0.315)  & (0.125)  & (0.155)  & (0.324)  & (0.091)  & (0.196) \\

$W_t\,\mathit{Market\mbox{-}to\mbox{-}book}$
& -0.005   & -0.003   & -0.002   & -0.005   & 0.001    & -0.012** \\
& (0.008)  & (0.003)  & (0.005)  & (0.008)  & (0.003)  & (0.005) \\

$W_t\,\mathit{Size}$
& -0.018   & -0.033*** & -0.018   & 0.024    & -0.018*  & -0.012 \\
& (0.023)  & (0.009)   & (0.016)  & (0.037)  & (0.010)  & (0.012) \\

$W_t\,\mathit{Profitability}$
& -0.174   & -0.084   & 0.117    & -0.469   & -0.166   & 0.089 \\
& (0.253)  & (0.125)  & (0.194)  & (0.513)  & (0.173)  & (0.279) \\

\midrule

Obs. (firms)
& 29 & 256 & 71 & 26 & 89 & 55 \\

\bottomrule
\end{tabular}%
}

\vspace{0.6em}
\begin{minipage}{1\textwidth}
\footnotesize
\textit{\textbf{Notes}:}
The table reports industry-specific Mean Group estimates for book leverage using broad SIC industry groupings. The dependent variable is book leverage. $W_{t-1}L.\mathit{Leverage}$ denotes network-weighted lagged book leverage constructed using the time-varying product-market weighting matrix. $W_t\,\mathit{Market\mbox{-}to\mbox{-}book}$,
$W_t\,\mathit{Size}$, and $W_t\,\mathit{Profitability}$ denote the corresponding network-weighted firm characteristics. The six reported industry groups contain 526 of the 549 firms in the baseline sample; firms belonging to the remaining SIC categories are omitted from the industry decomposition because the number of firms in those categories is too small for reliable Mean Group estimation. Standard errors, computed from the cross-sectional dispersion of the firm-specific coefficient estimates within each industry, are reported in parentheses. *, **, and *** denote statistical significance at the 10\%, 5\%, and 1\% levels, respectively.
\end{minipage}
\end{table}

\end{document}